\documentclass[preprint,pre,eqsecnum,longbibliography]{revtex4-1} 

\usepackage{natbib}
\usepackage{graphicx}
\usepackage{subfigure}
\usepackage{amssymb}
\usepackage{amsfonts}
\usepackage{amsmath}
\usepackage{amsbsy}
\usepackage{mathrsfs} 
\usepackage{color}
\usepackage{hyperref} 
\usepackage{soul}
\usepackage{lineno}
\usepackage{mymacros}

\begin{document} 

\begin{abstract}
In this Series, we study the weakly nonlinear dynamics of chemically active particles near the threshold for spontaneous motion. In this part, we focus on steady solutions and develop an `adjoint method' for deriving the nonlinear amplitude equation governing the particle's velocity, first assuming the canonical model in the literature of an isotropic chemically active particle and then considering general perturbations about that model. As in previous works, the amplitude equation is obtained from a solvability condition on the inhomogeneous problem at second order of a particle-scale weakly nonlinear expansion, the formulation of that problem involving asymptotic matching with a leading-order solution in a remote region where advection and diffusion are balanced. We develop a generalised solvability condition based on a Fredholm Alternative argument, which entails identifying the adjoint linear operator at the threshold and calculating its kernel. This circumvents the apparent need in earlier theories to solve the second-order inhomogeneous problem, resulting in considerable simplification and adding insight by making it possible to treat a wide range of perturbation scenarios on a common basis. To illustrate our approach, we derive and solve amplitude equations for a number of perturbation scenarios (external force and torque fields, non-uniform surface properties, first-order surface kinetics and bulk absorption), demonstrating that sufficiently near the threshold weak perturbations can appreciably modify and enrich the landscape of steady solutions. 
\end{abstract}

\title{Weakly nonlinear dynamics of a chemically active particle near the threshold for spontaneous motion. I. Adjoint method}

\author{Ory Schnitzer}
\affiliation{Department of Mathematics, Imperial College London, London SW7 2AZ, UK}


\maketitle
\section{Introduction}\label{sec:introduction}
Phoretic flows are surface-driven flows that result from local physico-chemical (chemical, thermal, electrical, etc.)~gradients within a fluid phase adjacent to a surface \cite{Anderson:89}. Such gradients may be either externally imposed or self-generated by an `active' surface. A prototypical example of the latter is the phoretic Janus particle, which self-propels as a consequence of its front and back sides having different properties \cite{Golestanian:05,Golestanian:07,Ebbens:14,Michelin:15,Popescu:16,Moran:17,Michelin:17}. Concomitantly, flow can modify physico-chemical gradients through advective transport. There is therefore mutual coupling between physico-chemical fields and liquid flow which under certain conditions results in unstable growth of perturbations. Together with the nonlinearity inherent to advection, such instabilities can lead to rich dynamics. A phenomenon that has received considerable attention in this context is symmetry breaking, where the physico-chemical and flow fields spontaneously form steady or unsteady structures possessing less symmetry than the governing equations. Examples include the formation of convection rolls in active channels \cite{Rubinstein:08,Game:17,Chen:21}, as well as the spontaneous dynamics exhibited by freely suspended and isotropic active particles \cite{Michelin:13,Michelin:14,Morozov:19,Chen:21,Kailasham:22}  and drops \cite{Rednikov:94,Schmitt:13,Izri:14,Suda:21,Hokmabad:21,Li:22,Hokmabad:21b,Michelin:22}. 

A canonical model for studying the dynamics of an isotropic active particle was introduced by Michelin \textit{et al.} \cite{Michelin:13}. It consists of a chemically active spherical particle that is freely suspended in an unbounded liquid solution. A single species of solute molecules is transported in the liquid bulk by diffusion and advection, approaching  an equilibrium concentration at infinity. The chemical activity of the particle is represented by a prescribed flux of solute molecules at the boundary of the particle and flow is driven at that boundary by diffusio-osmotic slip locally proportional to the surface gradient of the solute concentration. The prescribed solute flux and the slip coefficient are assumed uniform. This isotropic scenario allows for a steady state where both the fluid and particle are at rest and the solute distribution is spherically symmetric. It was shown by Michelin \textit{et al.}, however, that, depending on the signs of the solute flux and slip coefficient, the stationary-symmetric  state can be linearly unstable beyond a critical `intrinsic' P\'eclet number which quantifies the strength of advection relative to diffusion in the problem. At the threshold, the linear mode that becomes unstable describes steady rectilinear motion of the particle with arbitrary velocity and no rotation; the imaginary part of the growth rate vanishes there, implying a monotonic instability. Additional linear modes corresponding to higher wavenumbers become unstable at higher P\'eclet numbers, though these do not involve particle motion. 

Michelin \textit{et al.} \cite{Michelin:13} have also numerically simulated their canonical model as an initial value problem. These simulations show that, in an interval of P\'eclet numbers above the instability threshold, the particle approaches with time a steady state of spontaneous rectilinear motion in an arbitrary direction (determined in practice by initial conditions). As a function of the P\'eclet number, the speed of this spontaneous motion grows linearly away from the instability threshold, corresponding to an unconventional `singular' pitchfork bifurcation \cite{Farutin:21}. More recent simulations \cite{Chen:21,Kailasham:22} have revealed that at high P\'eclet numbers, at which the growth rates of higher-wavenumber linear modes dominate and the nonlinear spontaneous-motion states are expected to be unstable, the canonical particle exhibits complex and ultimately chaotic-like unsteady dynamics  \cite{Chen:21,Kailasham:22}. 

Several variations on the above canonical model of an isotropic active particle have been considered. These include variations to the chemical modeling that retain isotropy, such as accounting for first-order kinetics of the chemical reaction at the surface of the particle \cite{Michelin:14} or solute absorption in the bulk of the liquid solution \cite{Farutin:21}. Non-isotropic variations have also been considered, for example involving non-uniform surface properties \cite{Michelin:14,Saha:22}, geometric confinement \cite{Picella:22} or external force fields \cite{Yariv:17,Saha:21}. In the latter scenarios, there is a particular interest in how an imposed asymmetry influences the intrinsic spontaneous motion of the particle. The canonical model is also considered to be a `reference model' for active drops, which despite being more complex often exhibit qualitatively (and, in certain limits, quantitatively) similar dynamics \cite{Riazantsev:92,Morozov:19,Morozov:19b,Michelin:22}. There have been many studies of the influence of the environment on the spontaneous dynamics of active drops, for example involving external force fields \cite{Riazantsev:92,Rednikov:95}, surfactant transport \cite{Rednikov:94}, pair interactions \cite{Lippera:20,Lippera:20a}, cluster dynamics \cite{Hokmabad:21b}, bi-motility associated with viscosity variations   \cite{Hokmabad:21} and motion near boundaries \cite{Desai:21}. 
Incidentally, we note that several related  `toy models' have also been proposed, involving physically inconsistent simplifications of either the canonical model of an isotropic active particle or active-drop models. These include two- and three-dimensional `truncated' models where the concentration field (but not the flow field) is cut-off at some prescribed finite radius \cite{Hu:19,Farutin:21b,Hu:22,Li:22}, and point-particle models where advection in the vicinity of the particle is discarded \cite{Boniface:19,Farutin:21} or included in an \textit{ad hoc} manner \cite{Lippera:20b}.

Besides linear stability analysis and direct numerical simulations, several authors have applied weakly nonlinear analysis to describe various aspects of the spontaneous dynamics of chemically active particles and drops \cite{Riazantsev:92,Rednikov:94,Rednikov:95,Morozov:19,Morozov:19b,Lippera:20,Farutin:21,Saha:21,Li:22}. Most of these works have focused on the limit where the bifurcation parameter, the P\'eclet number in the canonical model, approaches its threshold value for instability and spontaneous motion. In that limit, weakly nonlinear analysis generally leads to a nonlinear `amplitude equation' governing the long-time dynamics, including the bifurcation of steady states, of the particle velocity vector. A useful feature of weakly nonlinear analysis is that it allows analytically treating perturbations \cite{Rednikov:95,Lippera:20,Saha:21}, which no matter how weak can still have an appreciable effect sufficiently close to the instability threshold. Some systems, such as deformable drops, involve multiple control parameters such that two linear modes can be tuned to lose stability at nearby P\'eclet numbers \cite{Ye:94,Morozov:19b,Farutin:21b}. In such scenarios, weakly nonlinear analysis allows deriving coupled amplitude equations governing the pair of modes \cite{Ye:94,Farutin:21b}.

For the most part, analysis near the onset of spontaneous  motion of an active particle or drop follows the standard paradigm of weakly nonlinear analysis near the threshold of a monotonic instability. Specifically, the first-order terms in the weakly nonlinear expansion correspond to a linearisation about the basic stationary state at the instability threshold. At this order, one finds a homogeneous, linear and quasi-static problem which is singular, having a family of homogeneous solutions representing spontaneous rectilinear motion of the particle with arbitrary velocity. Then at a higher order one finds an inhomogeneous version of that singular problem, whose solvability yields the requisite amplitude equation. A non-standard feature is that the weakly nonlinear expansion can be spatially nonuniform, holding in the vicinity of the particle or drop but not at large distances where advection and diffusion are comparable. Following classical analyses of forced heat advection at small P\'eclet numbers \cite{Acrivos:62}, this spatial non-uniformity can be resolved using the method of matched asymptotic expansions \cite{Hinch:book}, as done for steady states in \cite{Morozov:19,Morozov:19b,Saha:21}. 
This spatial non-uniformity of the asymptotics carries two important consequences. The first, explained by Farutin and Misbah \cite{Farutin:21}, is the unusual singular-pitchfork bifurcation of the spontaneous-motion steady states; related to this, the amplitude equation arises from a solvability condition at second, rather than third, order of the weakly nonlinear expansion. The second consequence, on which we elaborate below, has to do with the nature of the unsteady dynamics near the instability threshold. For some models of active particles and drops the spatial non-uniformity of the weakly nonlinear expansion is absent. That is the case, for example, for the truncated models mentioned previously \cite{Farutin:21b,Li:22}, models including strong absorption of solute in the bulk of the liquid solution  \cite{Farutin:21} and models of `reactive' active drops \cite{Michelin:22}, where solute transport is only considered in the interior of the drop and on its interface.  

While weakly nonlinear analysis is a natural avenue for studying the spontaneous dynamics of chemically active particles and drops, the existing literature suffers from two significant drawbacks which have so far limited the applicability of this approach. The first is concerned with the manner in which the solvability conditions that constitute the nonlinear amplitude equations have been derived. When faced with the question of solvability of an inhomogeneous linear problem, generally the preferred procedure is to apply the Fredholm Alternative to the forcing terms appearing in that problem \cite{Keener:18}; that  requires, however, knowledge of the relevant adjoint operator, which in the present context corresponds to the differential operator and auxiliary conditions adjoint to those at linear order of the weakly nonlinear expansion. This knowledge being absent, solvability conditions have to date been derived by solving the inhomogeneous problem in detail using separation of variables. This is a major technical complication, in particular because the linear operator (and its adjoint) is generally isotropic near the threshold for spontaneous motion --- with eigensolutions that are axisymmetric about an arbitrary axis --- whereas when allowing for general three-dimensional perturbations, or simply unsteadiness,  the inhomogeneous problem need not be. This perhaps explains why previous weakly nonlinear analyses have assumed \textit{a priori} that the motion of the particle is along a line and why genuinely three-dimensional problems, namely where such an assumption is not obvious or does not hold, have yet to be tackled. Furthermore, without employing the Fredholm Alternative the inhomogeneous problem needs to be solved for each physical scenario of interest, despite the relevant linear operator (and its adjoint) typically being common. The only related application of adjoint operators in the literature is a recent study by Farutin \textit{et al.} \cite{Farutin:21b}, who develop a pair of coupled amplitude equations starting from a two-dimensional truncated-domain model. Farutin \textit{et al.}, however, first project the governing equations onto Fourier modes and then separately seek the adjoint operator that is relevant to each mode; as we shall see, that is an unnecessary complication. 

The second drawback is that weakly nonlinear theories of active particles and drops near the threshold for spontaneous motion have so far been limited to steady or quasi-steady solutions, the only exceptions being models where the remote advection-diffusion region is absent \cite{Rednikov:95,Farutin:21b}. In conventional weakly nonlinear expansions about the threshold for a monotonic instability, accounting for unsteadiness amounts to identifying the long time scale associated with the closeness to the instability threshold --- often just that for which time derivatives first enter at the order of the weakly nonlinear analysis where solvability yields nonlinear amplitude equations. Unsteadiness is then manifested in the appearance of time derivatives in those equations, as is the case in \cite{Rednikov:95,Farutin:21b}. What appears to have been overlooked in the context of active particles and drops is that, when there is a remote advection-diffusion region, the relevant long time scale is actually associated with leading-order unsteadiness in that region. As a consequence, unsteadiness is manifested in the nonlinear amplitude equations as an integral over the history of the particle's motion, physically representing interaction of the particle with its self-generated concentration wake. 

In this Series, our goal is to address both of the above drawbacks. We will begin in this part by identifying the adjoint differential operator and auxiliary conditions relevant to the analysis of the canonical model of an isotropic chemically active particle near the threshold for spontaneous motion, as well as general perturbations about that model. To illustrate the application of our adjoint method, we shall, beyond recovering known results in the canonical case, consider the following four perturbation scenarios: (i) weak external force and torque fields, the force scenario being similar to \cite{Saha:21} but without making \textit{a priori} assumptions on symmetry; (ii) small arbitrary perturbations to the surface distributions of solute flux and the slip coefficient; (iii) inclusion of slow first-order kinetics at the surface of the particle, thus going beyond the model of a prescribed solute flux; and (iv) inclusion of weak absorption of solute in the liquid bulk. The first two perturbation scenarios are meant to demonstrate the efficacy of our approach in tackling genuinely three-dimensional problems. The latter two perturbation scenarios are included in order to demonstrate the less obvious applicability of our approach to perturbations involving different physico-chemical mechanisms. 

The adjoint method that we develop in this part is relevant to both steady and unsteady weakly nonlinear analyses. This is because the differential operator and auxiliary conditions encountered at linear order of the weakly nonlinear expansion (in the vicinity of the particle) are identical in both cases, owing to the long time scale associated with the closeness to the instability threshold. Nonetheless, the history effects associated with unsteadiness in the remote advection-diffusion region pose a formidable challenge that is rather separate from the solvability problem. We therefore choose to restrict this part to steady solutions and address unsteady problems involving history effects in a subsequent part. 

The structure of the paper is as follows. We begin in Sec.~\ref{sec:isotropic} by formulating the steady problem for the canonical model of an isotropic chemically active particle and, closely following the existing literature,  constructing the weakly nonlinear expansion for this case  in the limit as the P\'eclet number tends to its critical value. We terminate this section upon arriving at the inhomogeneous problem whose solvability yields the nonlinear amplitude equation. In this way, we set the stage for developing the adjoint method in Sec.~\ref{sec:adjoint}, ultimately leading to an explicit solvability condition for a generalised inhomogeneous problem which includes the canonical scenario as a special case. In Sec.~\ref{sec:perturbations}, we employ the generalised solvability condition to consider the perturbation scenarios mentioned above. We give concluding remarks in Sec.~\ref{sec:conclusions}, including a discussion of possible generalisations of our approach and a look ahead to the analysis of history-dependent unsteady dynamics in a subsequent part. 

\section{Weakly nonlinear expansion for the canonical model}\label{sec:isotropic}
\subsection{Physical problem}
We consider a spherical particle of radius $a_*$ that is freely suspended in an unbounded fluid of viscosity $\eta_*$. (An asterisk subscript indicates a dimensional quantity.) The fluid is a liquid solution in which a solute undergoes diffusion and advection, approaching an equilibrium concentration $\bar{c}_*$ at infinity. The solute also undergoes chemical reactions at the surface of the particle, represented by a prescribed flux  $j_*$ (positive into the liquid) that is assumed to be uniform. The latter flux is associated with the formation of a solute cloud in the vicinity of the particle, where the concentration perturbation from $\bar{c}_*$ is of order $c_* = a_*|j_*|/D_*$, $D_*$ being the solute diffusivity. 

Variations of solute concentration over the surface of the particle drive fluid flow through an effective diffusio-osmotic slip mechanism. In the standard continuum model, the local slip velocity is given by the local product of the concentration surface gradient and a scalar slip coefficient $b_*$, here assumed uniform. This slip mechanism, in conjunction with an order $c_*$ concentration perturbation varying on the particle scale, implies the intrinsic velocity scale $u_*=|j_*b_*|/D_*$. In turn, fluid flow can affect the distribution of solute through advection in the liquid bulk. The importance of solute advection relative to diffusion is quantified by the intrinsic P\'eclet number $\Pen = a_*u_*/D_*$. 

Since the particle is spherical and freely suspended, and the prescribed flux and slip coefficient uniform, there exists for all $\Pen$ a steady state where the particle and fluid are stationary and the solute cloud surrounding the particle is spherically symmetric. In the case $j_*b_*>0$, this stationary-symmetric state is known to be unstable for $\Pen>4$ \cite{Michelin:13}. Numerical simulations of initial-value problems \cite{Michelin:13,Chen:21,Kailasham:22} suggest a supercritical pitchfork bifurcation at the threshold into steady states of spontaneous rectilinear motion of the particle in an arbitrary direction, without rotation. Those numerical simulations also give the speed of that spontaneous motion as a function of $\Pen$, though only in a finite interval above the threshold where the dynamics indeed approach such spontaneous motion. More recently, several authors \cite{Morozov:19,Saha:21} have employed weakly nonlinear analysis to derive the local asymptotic behavior of the spontaneous speed near the bifurcation: 
\begin{equation}\label{local bifurcation}
{\text{spontaneous speed}} \sim {u_*}\frac{\Pen-4}{16} \quad \text{as} \quad \Pen\searrow 4, 
\end{equation}
assuming \textit{a priori}   that the motion is along a line and that the concentration and flow fields are axially symmetric. 

Our goal in this section is to review the derivation of the local bifurcation relation \eqref{local bifurcation} by a steady-state weakly nonlinear analysis in the limit $\Pen\to4$.  Specifically, we will carry through the analysis up to the final key step, in which \eqref{local bifurcation} arises from solvability of an  inhomogeneous problem at second order of the weakly nonlinear expansion. This will set the stage for the adjoint method developed in the subsequent section, which furnishes a solvability condition for a generalised inhomogeneous problem relevant to a wide range of physical scenarios, including but not limited to the perturbation scenarios considered in Sec.~\ref{sec:perturbations} and the extension in the next part to unsteady dynamics. 

We stress that the analysis in the present section closely follows \cite{Morozov:19,Saha:21} except that the solvability condition is quoted based on the adjoint method to be developed instead of a detailed solution of the inhomogeneous problem. We also avoid \textit{a priori} assumptions on symmetry, a generality which will be crucial in Sec.~\ref{sec:perturbations} when considering non-isotropic perturbations and in the next part when considering unsteady dynamics. 

As mentioned above, spontaneous motion occurs only if $j_*b_*>0$. We shall assume, without loss of generality, that $j_*$ and $b_*$ are both positive. 

\subsection{Dimensionless formulation} \label{ssec:formu}
We adopt a dimensionless formulation where lengths are normalised by $a_*$, concentration by $c_*$, velocities by $u_*$, stresses by $\eta_*u_*/a_*$, forces by $\eta_*u_*a_*$ and torques by $\eta_*u_*a_*^2$. With these conventions, the position vector measured from the center of the particle is denoted by $\br=r\be_r$, with $r=|\br|$ and $\be_r$ a radial unit vector such that the surface of the sphere is at $r=1$ and the fluid domain is $r>1$; the solute concentration is denoted by $c$; the fluid velocity field, in a `co-moving' frame of reference that moves with the particle's centroid without rotating, is denoted by $\bu$; the pressure field associated with that flow is denoted by $p$; the velocity of the particle centroid is denoted by $\bU$; and the particle angular velocity is denoted by $\boldsymbol{\Omega}$. We look for steady-state solutions such that the particle velocities $\bU$ and $\boldsymbol{\Omega}$ are constant vectors and the fields $c,\bu$ and $p$ are constant in time in the co-moving frame. 

The complete formulation is composed of two problems that are mutually coupled. The first can be thought of as governing the concentration field $c$. It consists of the steady advection-diffusion equation
\begin{equation}\label{c eq}
\nabla^2c -\Pen\,\bu\bcdot\bnabla c=0;
\end{equation}
the boundary condition 
\begin{equation}\label{c bc}
\pd{c}{r}=-1 \quad \text{at} \quad r=1;
\end{equation}
and the decay condition
\begin{equation}\label{c far}
c\to0 \quad \text{as} \quad r\to\infty.
\end{equation}

The second problem can be thought of as governing the flow field $\bu$ and pressure field $p$, which without loss of generality is taken to decay at infinity. The flow problem consists of the Stokes equations 
\refstepcounter{equation}
$$
\label{u eqs}
\bnabla \bcdot \bu  = 0, \quad \bnabla\bcdot\boldsymbol{\sigma}=\bzero,
\eqno{(\theequation a,\!b)}
$$
in which 
\begin{equation}\label{stress}
\boldsymbol{\sigma}=-p\tI + \bnabla\bu + (\bnabla\bu)^\dagger
\end{equation} 
is the stress tensor, $\tI$ being the identity tensor and $\dagger$ denoting the tensor transpose; 
the boundary condition
\begin{equation}\label{u bc}
\bu=\bnabla_s c +\mathbf{\Omega}\times \br \quad \text{at} \quad r=1,
\end{equation}
wherein $\bnabla_s$ denotes the surface-gradient operator (see Appendix \ref{app:surfaceop}); 
the far-field condition
\begin{equation}\label{u far}
\bu \to -\mathbf{U} \quad \text{as} \quad r\to\infty;
\end{equation}
and the integral conditions
\refstepcounter{equation}
$$
\label{force torque balances}
\bF= \bzero, \quad \bT=\bzero,
\eqno{(\theequation a,\!b)}
$$
where the force $\bF$ and torque $\bT$ are defined as 
\refstepcounter{equation}
$$
\label{force torque}
\bF= \oint_{r=1} \mathrm{d}A\, \be_r\bcdot \boldsymbol{\sigma}, \quad \bT= \oint_{r=1} \mathrm{d}A\, \br\times (\be_r\bcdot \boldsymbol{\sigma}),
\eqno{(\theequation a,\!b)}
$$
wherein $\mathrm{d}A$ is an infinitesimal area element.

\subsection{Asymptotic expansions}
As previously mentioned, there exists for arbitrary $\Pen$ a solution where the particle is stationary, the fluid is at rest, and the concentration field is spherically symmetric. Indeed, the flow problem is then trivially satisfied while the advection-diffusion equation \eqref{c eq} reduces to Laplace's equation. Together with the boundary condition \eqref{c bc} and far-field decay \eqref{c far}, we find the concentration field
\begin{equation}\label{c0}
c_0=\frac{1}{r}.
\end{equation}
In what follows we look for additional steady solutions as $\Pen\to4$, the limiting value being known to be the threshold above which the stationary-symmetric state is linearly unstable. To be consistent with the perturbation scenarios considered later in the paper, it is convenient to write 
\begin{equation}\label{chi def}
\Pen=4 + \epsilon \chi
\end{equation}
and then consider the limit $0<\epsilon\ll1$, with $\chi$ independent of $\epsilon$. In this  section, only the product $\epsilon \chi$ has meaning so $\chi$ can, in principle, be chosen as plus or minus unity without loss of generality.  

We attempt an expansion of the concentration field in powers of $\epsilon$:  
\begin{equation}\label{c expansion}
c(\br;\epsilon) \sim c_0(r)+\epsilon c_1(\br)+\epsilon^2 c_2(\br) \quad \text{as} \quad \epsilon\searrow0,
\end{equation}
with $c_0$ provided by \eqref{c0}. Similarly, we assume that the velocity and pressure fields are expanded as 
\refstepcounter{equation}
$$
\label{u p expansions}
\bu(\br;\epsilon) \sim \epsilon\bu_{1}(\br) + \epsilon^2 \bu_2(\br), \quad p(\br;\epsilon)\sim \epsilon p_{1}(\br)+\epsilon^2 p_2(\br) \quad \text{as} \quad \epsilon\searrow0.
\eqno{(\theequation a,\!b)}
$$
Quantities associated with the flow field are expanded similarly to \eqref{u p expansions}. These include the stress tensor $\boldsymbol{\sigma}$, force $\bF$ and torque $\mathbf{T}$, as well as the particle velocities $\bU$ and $\boldsymbol{\Omega}$. 

Expansion \eqref{c expansion} for the concentration, with $\br$ fixed, represents a  `particle region' corresponding to order unity distances from the particle. Allowing $\br$ to vary, this expansion fails at large, order $1/\epsilon$, distances from the particle, corresponding to a `remote region' where advection and diffusion are comparable. Indeed, if the concentration field decays like $1/r$ and the velocity field scales like $\epsilon$ and approaches a uniform stream, the advection and diffusion terms in \eqref{c eq} are of order $\epsilon/r^2$ and $1/r^3$, respectively, for large $r$. Accordingly, it is necessary to supplement the particle-region expansion \eqref{c expansion} with a remote-region expansion in which $\epsilon \br$, rather than $\br$, is held fixed. Asymptotic matching \cite{Hinch:book} of the particle and remote regions will be employed in order to derive far-field conditions as $r\to\infty$ at subsequent orders of the particle-scale concentration expansion, effectively replacing the decay condition \eqref{c far}. We will find that the assumed decay of the leading term $c_0$ in the particle-scale expansion is consistent with asymptotic matching, despite that term not holding as a leading-order approximation in the remote region. 

\subsection{Remote region}\label{ssec:remote}
To analyze the remote region, we define  the strained position vector $\tilde{\br}=\epsilon \br$, with $\tilde{r}=|\tilde{\br}|=\epsilon r$,  make the change of variables $\tilde{c}(\tilde{\br};\epsilon)=c(\br;\epsilon)$ and consider the limit $\epsilon\searrow0$ with $\tilde{r}>0$ fixed. The $1/r$ decay of the particle-region concentration, along with the scaling of the remote region, suggests that $\tilde{c}$ is of order $\epsilon$ in that region. We therefore pose the expansion
\begin{equation}\label{remote expansion}
\tilde{c}(\tilde{\br};\epsilon)\sim \epsilon\tilde{c}_{1}(\tilde{\br}) \quad \text{as} \quad \epsilon\searrow0. 
\end{equation}
Unlike the concentration field, the expansion for the flow field is clearly uniformly valid for $r>1$. Hence the far-field condition \eqref{u far} and the expansion for $\bU$ (cf.~\eqref{u p expansions}) together imply that $\bu\sim -\epsilon \bU_1$ holds in the remote region. It then follows from \eqref{c eq} that $\tilde{c}_1$ satisfies an advection-diffusion equation with a uniform flow,
\begin{equation}\label{remote eq} \tilde{\nabla}^2\tilde{c}_1+4\bU_1\bcdot\tilde{\bnabla}\tilde{c}_1=0,
\end{equation}
in which $\tilde{\bnabla}$ is the gradient operator with respect to $\tilde{\br}$. Furthermore, the decay condition
\begin{equation}\label{remote far}
\tilde{c}_1\to0\quad\text{as} \quad \tilde{r}\to\infty
\end{equation}
follows from \eqref{c far}, while the singular boundary condition 
\begin{equation}\label{remote matching}
\tilde{c}_1\sim \frac{1}{\tilde{r}}\quad \text{as} \quad \tilde{r}\to0
\end{equation}
follows from asymptotic matching between the one-term remote-region expansion and the particle-region expansion taken to leading order. The solution to \eqref{remote eq}--\eqref{remote matching} is \cite{Acrivos:62}
\begin{equation}\label{remote sol}
\tilde{c}_1=\frac{1}{\tilde{r}}\exp\left\{-2{\bU}_1\bcdot \tilde{\br}-2|{\bU}_1|\tilde{r}\right\}.
\end{equation}
Higher-order matching between the one-term remote-region expansion and the particle-region expansion to orders $\epsilon$ and $\epsilon^2$ will be seen to provide sufficient information on the far-field behavior of the particle-region fields $c_1$ and $c_2$, respectively. 

\subsection{Homogeneous problem}
\label{ssec:homogeneous}
Returning to the particle region, we consider the homogeneous problem governing the order-$\epsilon$ concentration and flow field. The concentration problem \eqref{c eq}--\eqref{c far} gives the coupled advection-diffusion equation
\begin{equation}\label{c1 eq}
\nabla^2c_{1}-4\bu_{1}\bcdot\bnabla \frac{1}{r}=0
\end{equation}
and the boundary condition
\begin{equation}\label{c1 bc}
\pd{c_{1}}{r}=0 \quad \text{at} \quad r=1,
\end{equation}
while asymptotic matching with the remote region implies the far-field condition
\begin{equation}\label{c1 far}
c_1= -2\bU_1\bcdot\be_r -2|\bU_1| +o(1) \quad \text{as} \quad r\to\infty.
\end{equation}
The flow problem \eqref{u eqs}--\eqref{force torque balances} gives the Stokes equations
\refstepcounter{equation}
$$
\label{stokes 1}
\bnabla \bcdot \bu_{1}  = 0, \quad \bnabla\bcdot\boldsymbol{\sigma}_1=\bzero; 
\eqno{(\theequation a,\!b)}
$$
the boundary condition
\begin{equation}\label{u1 bc}
\bu_{1}=\bnabla_s c_{1}+\boldsymbol{\Omega}_1\times\br \quad \text{at} \quad r=1;
\end{equation}
the far-field condition 
\begin{equation}\label{u1 far}
\bu_{1}\to -\bU_1 \quad \text{as} \quad r\to\infty; 
\end{equation}
and the force and torque balances 
\refstepcounter{equation}
$$
\label{force torque balances 1}
\bF_{1}=\bzero, \quad \bf{T}_{1}=\bzero.
\eqno{(\theequation a,\!b)}
$$

\subsection{Spontaneous motion}
\label{ssec:spontaneous}
The order-$\epsilon$ problem is homogeneous; it is also linear, except for the nonlinear constant term in the matching condition \eqref{c1 far}, which merely determines a uniform reference value for $c_1$. Despite being homogeneous, the order-$\epsilon$ problem possesses a family of non-trivial solutions corresponding to steady rectilinear motion of the particle with an arbitrary velocity vector $\bU_1$ and without rotation, i.e., $\boldsymbol{\Omega}_1=\bzero$. The existence of these solutions is no coincidence but an expected consequence of perturbing about the threshold value of the P\'eclet number. These homogeneous solutions have been calculated in \cite{Morozov:19} using separation of variables in spherical coordinates. We here rewrite these solutions as
\refstepcounter{equation}
$$
\label{homogeneous sol}
c_1 = -2|\bU_1|+c_L(\br;\bU_1) , \quad \bu_{1}=\bu_L(\br;\bU_1), \quad p_{1}=p_L(\br;\bU_1), 
\eqno{(\theequation a\!\!-\!\!c)}
$$
where
\refstepcounter{equation}
$$
\label{linear sol}
c_L(\br;\bU_1) =  4\bU_1\bcdot g(r)\br , \quad \bu_L(\br;\bU_1)=-\bU_1+\frac{1}{2}\bU_1\bcdot\bnabla\bnabla\frac{1}{r}, \quad p_L(\br;\bU_1)=0, 
\eqno{(\theequation a\!\!-\!\!c)}
$$
are linear in $\bU_1$, with $g(r)$ being the radial function
\begin{equation}\label{radial function}
g(r)=\frac{3}{8r^3}-\frac{1}{2r}-\frac{1}{4r^4}.
\end{equation}

\subsection{Inhomogeneous problem}
\label{ssec:inhomogeneous}
We next consider the inhomogeneous problem governing the order-$\epsilon^2$ concentration and flow fields, whose solvability will be seen to restrict the particle velocity $\bU_1$. 

The concentration problem \eqref{c eq}--\eqref{c far} gives the inhomogeneous coupled advection-diffusion equation
\begin{equation}\label{c eq forced}
\nabla^2c_2-4\bu_2\bcdot\bnabla\frac{1}{r}=4\bu_1\bcdot\bnabla c_1+\chi\bu_1\bcdot\bnabla\frac{1}{r}
\end{equation}
and the boundary condition 
\begin{equation}\label{c bc forced}
\pd{c_2}{r}=0 \quad \text{at} \quad r=1.
\end{equation}
The inhomogeneous far-field condition
\begin{multline}\label{c far forced}
c_2 = r\left\{2\left(\tI+\be_r\be_r\right)\boldsymbol{:}\bU_1\bU_1+4\be_r\bcdot\bU_1|\bU_1|\right\} \\ -\frac{\chi}{2}\bU_1\bcdot\be_r - 2\bU_2\bcdot\be_r  + \text{const.} + o(1) \quad \text{as} \quad r\to\infty
\end{multline}
is derived as follows. First, matching the particle-scale expansion to order $\epsilon^2$ with the remote-scale expansion to order $\epsilon$ determines the order-$r$ terms. Second, a local analysis of \eqref{c eq forced} as $r\to\infty$,  using \eqref{u1 far}, \eqref{homogeneous sol} and \eqref{u far forced}, is consistent with those order-$r$ terms and further implies the form of the order-unity terms. In particular, the constant term in \eqref{c far forced}, which sets the uniform reference value of $c_2$, could be derived by higher-order matching involving a first correction in the remote region. As we shall see, however, only the order-$r$ forcing terms are relevant to the matter of solvability. 

The flow problem \eqref{u eqs}--\eqref{force torque balances} gives  the Stokes equations 
\refstepcounter{equation}
$$
\label{u eqs forced}
\bnabla\bcdot\bu_2 = 0, \quad \bnabla\bcdot\boldsymbol{\sigma}_2=\bzero;
\eqno{(\theequation a,\!b)}
$$
the boundary condition
\begin{equation}\label{u bc forced}
\bu_2=\bnabla_s c_2+\boldsymbol{\Omega}_2\times\br \quad \text{at} \quad r=1;
\end{equation}
the far-field condition
\begin{equation}\label{u far forced}
\bu_2\to -\bU_2 \quad \text{as} \quad r\to\infty;
\end{equation}
and the integral constraints 
\refstepcounter{equation}
$$
\label{F T balances forced}
\bF_2=\bzero, \quad \bT_2=\bzero.
\eqno{(\theequation a,\!b)}
$$

\subsection{Solvability condition yields nonlinear amplitude equation}
The order-$\epsilon^2$ problem is an inhomogeneous version of the order-$\epsilon$ homogeneous problem. Since the latter problem is singular, we anticipate that the inhomogeneous order-$\epsilon^2$ problem is solvable only under a certain condition on the forcing terms, suggesting a relation between the bifurcation parameter $\chi$ and the particle velocity $\bU_1$. In order to derive this solvability condition, it is not necessary to construct detailed solutions to the inhomogeneous problem. Rather, in the next section we shall derive a solvability condition for a generalised inhomogeneous problem with the help of an adjoint linear operator.  The specific solvability condition that is relevant here, which is obtained in Sec.~\ref{ssec:basicsolvability}, reads
\begin{equation}\label{solvability isotropic}
\bU_1\left(16|\bU_1|-\chi\right)=\bzero.
\end{equation}

The solvability condition \eqref{solvability isotropic} constitutes the requisite nonlinear amplitude equation, which in the present steady formulation serves as a local bifurcation relation. For arbitrary $\chi$ there is always the trivial solution, $\bU_1=\bzero$, which is consistent with the stationary-symmetric state. For $\chi>0$, there are also non-trivial solutions corresponding to steady rectilinear motion without rotation, in an arbitrary direction and with speed 
\begin{equation}\label{abs U1 isotropic}
|\bU_1|=\frac{\chi}{16}.
\end{equation}
Rewriting \eqref{abs U1 isotropic} in dimensional notation implies the local bifurcation relation \eqref{local bifurcation} stated at the beginning of this section. 
The singular-pitchfork bifurcation implied by \eqref{solvability basic} is schematically depicted in Fig.~\ref{fig:force}a, recalling that in the present scenario it suffices to consider $\chi=\pm1$. As indicated in the figure, the axisymmetric linear-stability analysis in \cite{Michelin:13} and three-dimensional numerical simulations in \cite{Michelin:13,Kailasham:22,Chen:21} suggest that, for $\chi>0$,  the trivial solution loses stability in favour of the non-trivial spontaneous-motion states. 

\section{Adjoint method}\label{sec:adjoint}
\subsection{Linear operator and the direct problem}\label{ssec:homogeneousR}
To develop the adjoint method we start by formally defining the `matrix' differential operator 
\begin{equation}\label{L def}
\mathcal{L} =\left(\begin{array}{cc}\nabla^2& -4\bnabla\left(\frac{1}{r}\right)\bcdot \\0 & \bnabla\bcdot\boldsymbol{S}\end{array}\right),
\end{equation}
which acts on concentration-flow `pairs', namely  `column vectors' $\underset{\sim}{\psi}=(c\,\,\bu)^T$ in which $c$ is a concentration field and $\bu$ an incompressible flow field that has associated with it a pressure field $p$. In \eqref{L def}, $\boldsymbol{S}$ represents the stress-tensor operator
\begin{equation}
\boldsymbol{S}[\bu]=-p\tI+\bnabla\bu + (\bnabla\bu)^\dagger
\end{equation}
and the components of $\mathcal{L}$ act on the components of $\underset{\sim}{\psi}$ analogously to standard matrix multiplication. We also define a set of auxiliary conditions, involving particle velocities $\bU$ and $\boldsymbol{\Omega}$, and we shall say that a pair satisfying these conditions is included in the `natural domain' of $\mathcal{L}$. These auxiliary conditions consist of the boundary conditions 
\refstepcounter{equation}
$$
\label{bcs linear}
\pd{c}{r}=0, \quad \bu= \bnabla_s c + \boldsymbol{\Omega}\times \br \quad \text{at} \quad r=1;
\eqno{(\theequation a,\!b)}
$$
the far-field conditions
\refstepcounter{equation}
$$
\label{far linear}
c  = -2\bU\bcdot\be_r+\mathcal{E}, \quad \bu \to -\bU \quad \text{as} \quad r\to\infty,
\eqno{(\theequation a,\!b)}
$$
where $\mathcal{E}=o(1)$ and $\bnabla\mathcal{E}=o(1)$ as $r\to\infty$; 
and the integral constraints
\refstepcounter{equation}
$$
\label{F T balances linear}
{\bF}=\bzero, \quad \bT=\bzero,
\eqno{(\theequation a,\!b)}
$$
where the torque $\bT$ is defined as in (\ref{force torque}b) while, henceforth, we adopt the following definition for the force $\bF$:
\begin{equation}
\label{F def new}
{\bF}=\lim_{R\to\infty}\oint_{r=R}\mathrm{d}A\,\be_r\bcdot \boldsymbol{\sigma},
\end{equation}
wherein $\boldsymbol{\sigma}=\boldsymbol{S}[\bu]$. The modified force definition \eqref{F def new} agrees with the conventional definition (\ref{force torque}a) in the case where the stress is divergence-free but is otherwise more specific.

With the above definitions, we define the `direct' problem 
\begin{equation}\label{homogeneous reformulation}
\mathcal{L} \underset{\sim}{{\psi}}=\underset{\sim}{0},
\end{equation}
where $\underset{\sim}{0}$ is the zero pair and $\underset{\sim}{\psi}$ is restricted to the natural domain of $\mathcal{L}$. This direct problem is equivalent to the order-$\epsilon$ homogeneous problem of Sec.~\ref{ssec:homogeneous}, except for the different uniform reference value of the concentration field. 
Written for $\underset{\sim}{{\psi}}=(c\,\,\bu)^T$ and associated particle velocities $\bU$ and $\boldsymbol{\Omega}$, the direct problem possesses the family of homogeneous solutions $c=c_L(\br;\bU)$ and $\bu=\bu_L(\br;\bU)$, with $\bU$ arbitrary and $\boldsymbol{\Omega}=\bzero$ (cf.~\eqref{linear sol}).

\subsection{Adjoint operator}\label{ssec:adjointoperator}
We now introduce another matrix differential operator, 
\begin{equation}\label{L star def}
\mathcal{L}^* =\left(\begin{array}{cc}\nabla^2& 0 \\ -4\bnabla\left(\frac{1}{r}\right) & \bnabla\bcdot\boldsymbol{S}\end{array}\right),
\end{equation}
which is essentially the transpose of $\mathcal{L}$, except for the omission of the `dot' operator from the off-diagonal term (implied by that term now operating on a scalar to give a vector). Similar to $\mathcal{L}$, the operator $\mathcal{L}^*$ acts on concentration-flow pairs where the flow is incompressible and has associated with it a pressure field. We also define a set of adjoint auxiliary conditions, and we shall say that pairs satisfying these conditions are included in the natural domain of $\mathcal{L}^*$. We specify these  adjoint auxiliary conditions considering a primed pair $(c'\,\,\bu')^T$, with associated pressure $p'$, stress tensor $\boldsymbol{\sigma}'=\boldsymbol{S}[\bu']$ and particle velocities $\bU'$ and $\boldsymbol{\Omega}'$. The adjoint auxiliary conditions consist of the boundary conditions
\refstepcounter{equation}
$$
\label{bcs A}
\pd{c'}{r}=\bnabla_s\bcdot \left[(\tI-\be_r\be_r)\bcdot (\be_r\bcdot\boldsymbol{\sigma}')\right], \quad \bu'= \boldsymbol{\Omega}'\times \br \quad \text{at} \quad r=1,
\eqno{(\theequation a,\!b)}
$$
where $\bnabla_s\bcdot$ is the surface-divergence operator (see Appendix \ref{app:surfaceop}); 
the far-field conditions
\refstepcounter{equation}
$$
\label{far A}
c' = \mathcal{E}', \quad \bu'\to -\bU' \quad \text{as} \quad r\to\infty,
\eqno{(\theequation a,\!b)}
$$
where $\mathcal{E}'=\mathcal{O}\left({1}/{r^2}\right)$ and $\bnabla\mathcal{E}'=\mathcal{O}\left({1}/{r^3}\right)$ as $r\to\infty$; 
and the integral constraints
\refstepcounter{equation}
$$
\label{F T balances A}
\bF'=\bzero, \quad \bT'=\bzero.
\eqno{(\theequation a,\!b)}
$$

We shall now verify that the operator $\mathcal{L}^*$ with its natural domain is formally the adjoint of the operator $\mathcal{L}$ with its natural domain, with respect to the inner product 
\begin{equation}
\langle \underset{\sim}{\psi} ,\underset{\sim}{\varphi} \rangle = \lim_{R\to\infty} \int_{\mathcal{D}_R}\mathrm{d}V\,\underset{\sim}{\psi}^T \cdot \underset{\sim}{\varphi},
\end{equation}
where $\underset{\sim}{\psi}$ and $\underset{\sim}{\varphi}$ denote concentration--flow pairs; $\mathcal{D}_R$ is the domain $1<r<R$;  the dot product inside the integral is defined as the generalised scalar product given by the sum of the product of concentrations and the scalar product of flow fields; and $\mathrm{d}V$ denotes an infinitesimal volume element. To confirm the adjoint property, we need to verify that the difference
\begin{equation}\label{J def}
J(\underset{\sim}{\psi},\underset{\sim}{\varphi})=\langle \mathcal{L} \underset{\sim}{\psi},\underset{\sim}{\varphi}\rangle -\langle \underset{\sim}{\psi},\mathcal{L}^*\underset{\sim}{\varphi}\rangle
\end{equation}
vanishes trivially for any pair $\underset{\sim}{\psi}=(c\,\,\bu)^T$ in the natural domain of $\mathcal{L}$ and pair $\underset{\sim}{\varphi}=(c'\,\,\bu')^T$ in the natural domain of $\mathcal{L}^*$. It is this requirement that has guided us in defining the differential operator $\mathcal{L}^*$ and its natural domain as we have. 

We can write \eqref{J def} explicitly as
\begin{equation}\label{J general volume}
J(\underset{\sim}{\psi},\underset{\sim}{\varphi})=\lim_{R\to\infty}\left\{\int_{\mathcal{D}_R}\mathrm{d}V\,\left(c'\nabla^2c-c\nabla^2c'\right)+\int_{\mathcal{D}_R}\mathrm{d}V\,\left(\bu'\bcdot\bnabla\bcdot\boldsymbol{\sigma}-\bu\bcdot\bnabla\bcdot\boldsymbol{\sigma}'\right)\right\},
\end{equation}
where we note that the terms involving the off-diagonal coupling terms in \eqref{L def} and \eqref{L star def} have cancelled out. The volume integrals in \eqref{J general volume} can be transformed into surface integrals. For the first integral, we use the divergence theorem applied to $c'\nabla c-c\nabla c'$ (i.e., we use Green's second identity). Analogously,  for the second integral we use the divergence theorem applied to $\boldsymbol{\sigma}\bcdot\bu'-\boldsymbol{\sigma}'\bcdot\bu$, the definitions of the stress tensors and the fact that the flows are incompressible (as in the proof of the Lorenz Reciprocal Theorem for Stokes flow \cite{Happel:book}). Overall, we find
\begin{equation}\label{J general surface}
J(\underset{\sim}{\psi},\underset{\sim}{\varphi}) = \lim_{R\to\infty}\left\{\oint_{\partial\mathcal{D}_R}\mathrm{d}A\,\left(c'\pd{c}{n}-c\pd{c'}{n}\right)+\oint_{\partial\mathcal{D}_R} \mathrm{d}A\,\left[\bu'\bcdot (\bn\bcdot\boldsymbol{\sigma})-\bu\bcdot\left(\bn\bcdot\boldsymbol{\sigma}'\right)\right]\right\},
\end{equation}
where $\partial{\mathcal{D}}_R$ is the boundary of $\mathcal{D}_R$ and $\bn$ is the normal unit vector in the direction outward from $\mathcal{D}_R$, with $\partial/\partial{n}=\bn\bcdot\bnabla$. 

The boundary $\partial\mathcal{D}_R$ is composed of the surface of the unit sphere and the surface of the sphere $r=R$. Separating these contributions and applying the natural boundary conditions at $r=1$ (cf.~\eqref{bcs linear} and \eqref{bcs A}), we find
\begin{multline}
\label{three lines}
J(\underset{\sim}{\psi},\underset{\sim}{\varphi})=\oint_{r=1} \mathrm{d}A\,\left\{c\bnabla_s\bcdot \left[(\tI-\be_r\be_r)\bcdot (\be_r\bcdot\boldsymbol{\sigma}')\right]+\bnabla_s c \bcdot (\be_r\bcdot\boldsymbol{\sigma}')\right\}\\
-\oint_{r=1} \mathrm{d}A\,(\boldsymbol{\Omega}'\times\br)\bcdot (\be_r\bcdot\boldsymbol{\sigma})
+\oint_{r=1} \mathrm{d}A\,(\boldsymbol{\Omega}\times\br)\bcdot (\be_r\bcdot\boldsymbol{\sigma}')
\\
+\lim_{R\to\infty}\left\{
\oint_{r=R}\mathrm{d}A\,\left(c'\pd{c}{r}-c\pd{c'}{r}\right)
+\oint_{r=R} \mathrm{d}A\,\left[\bu'\bcdot (\be_r\bcdot\boldsymbol{\sigma})-\bu\bcdot\left(\be_r\bcdot\boldsymbol{\sigma}'\right)\right]\right\}.
\end{multline}
The first line of \eqref{three lines} vanishes by `integration by parts' over the surface of the unit sphere. Indeed, we have that
\begin{equation}\label{manipulation}
c\bnabla_s\bcdot \left[(\tI-\be_r\be_r)\bcdot (\be_r\bcdot\boldsymbol{\sigma}')\right]+\bnabla_s c \bcdot (\be_r\bcdot\boldsymbol{\sigma}')=
\bnabla_s\bcdot\left[c(\tI-\be_r\be_r)\bcdot (\be_r\bcdot\boldsymbol{\sigma}')\right],
\end{equation}
and the integral over a closed surface of the surface divergence of a tangential vector field trivially vanishes (see Appendix \ref{app:surfaceop}). 
The second line can be seen to vanish using the triple-product rule and the fact that the torques $\bT$ and $\bT'$ vanish for flow fields in the natural domains (cf.~(\ref{F T balances linear}b) and (\ref{F T balances A}b)).
It remains to show that the last line of \eqref{three lines} vanishes. The limit of the concentration integrals is readily seen to vanish on account of the natural far-field  conditions (\ref{far linear}a) and (\ref{far A}a). The stress integrals also vanish, since 
\begin{multline}\label{last line u}
\lim_{R\to\infty}\oint_{r=R} \mathrm{d}A\,\left[\bu'\bcdot (\be_r\bcdot\boldsymbol{\sigma})-\bu\bcdot\left(\be_r\bcdot\boldsymbol{\sigma}'\right)\right]
\\=-\mathbf{U}'\bcdot \lim_{R\to\infty}\oint_{r=R} \mathrm{d}A\,(\be_r\bcdot\boldsymbol{\sigma})
+\bU\bcdot\lim_{R\to\infty}\oint_{r=R} \mathrm{d}A\,(\be_r\bcdot\boldsymbol{\sigma}')=-\bU'\bcdot\bF+\bU\bcdot\bF'=0,
\end{multline}
where we have used the natural far-field conditions and force constrains. We note that this last step rationalises our force definition \eqref{F def new}.

\subsection{Adjoint problem and adjoint spontaneous motion}
\label{ssec:adjointproblem}
Consider now the `adjoint problem' for a concentration-flow pair $\underset{\sim}{\psi'}=(c'\,\,\bu')^T$, say, which is defined by 
\begin{equation}\label{adjoint problem}
\mathcal{L}^*\underset{\sim}{\psi'}=0,
\end{equation}
with $\underset{\sim}{\psi'}$ restricted to the natural domain of $\mathcal{L}^*$. 

In this adjoint problem, the coupling between concentration and flow is physically transposed. Thus, while in the direct problem stress is divergence-free and the divergence of concentration flux is proportional to the flow field, in the adjoint problem it is the concentration flux that is divergence-free whereas the divergence of stress is proportional to the concentration. Explicitly, \eqref{adjoint problem} gives Laplace's equation 
\begin{equation}\label{c eq A}
\nabla^2c'=0
\end{equation}
and, together with the condition that $\bu'$ is incompressible, the concentration-coupled Stokes equations
\refstepcounter{equation}
$$
\label{u eqs A}
\bnabla\bcdot\bu'=0, \quad -4 c'\bnabla\frac{1}{r}+\bnabla\bcdot\boldsymbol{\sigma}'=\bzero.
\eqno{(\theequation a,\!b)}
$$
Similarly, while in the direct problem the normal surface flux of concentration vanishes and the surface slip velocity is proportional to the surface gradient of concentration (cf.~\ref{bcs linear}), in the adjoint homogeneous problem it is the surface slip velocity that vanishes whereas the normal surface flux of concentration is proportional to the surface divergence of the tangential traction (cf.~\ref{bcs A}). 

The adjoint homogeneous problem, like the homogeneous problem, possesses non-trivial solutions that represent spontaneous rectilinear motion with arbitrary particle velocity $\bU'$, without rotation, i.e., $\boldsymbol{\Omega}'=\bzero$. In Appendix \ref{app:spontaneousA}, we find these solutions as
\refstepcounter{equation} 
$$
\label{sol associated}
c'=-\frac{3\mathbf{U}'\bcdot \br}{r^3}, \quad \bu'=\bU'\bcdot\left[\left(-1+\frac{1}{r^3}\right)\tI+3\br\br\left(\frac{1}{r^4}-\frac{1}{r^5}\right)\right], \quad p'=\frac{6\mathbf{U}'\bcdot\br}{r^4},
\eqno{(\theequation a\!\!-\!\!c)}.
$$
A property of these solutions that will be useful later is the projection of the traction tangent to the sphere: 
\begin{equation}\label{stress result}
(\tI-\be_r\be_r)\bcdot (\be_r\bcdot\boldsymbol{\sigma}')=-3\bU'\bcdot (\tI-\be_r\be_r) \quad \text{at} \quad r=1. 
\end{equation}

\subsection{Necessary condition for solvability of a generalised inhomogeneous problem}
Consider now an inhomogeneous problem of the form
\begin{equation}\label{inhomogeneous} 
\mathcal{L}\underset{\sim}{\psi}=\underset{\sim}{f},
\end{equation}
with $\underset{\sim}{f}$ a prescribed scalar-vector pair and $\underset{\sim}{\psi}$ not necessarily in the natural domain of $\mathcal{L}$. This problem is inhomogeneous in both the partial differential equations implied by \eqref{inhomogeneous} and the auxiliary conditions satisfied by $\underset{\sim}{\psi}$, to be specified below. Let $\underset{\sim}{\varphi}$ be any pair in the kernel of $\mathcal{L}^*$ that is in the natural domain of $\mathcal{L}^*$, i.e., any solution of the adjoint homogeneous problem. Since $\underset{\sim}{\psi}$ is not necessarily in the natural domain of $\mathcal{L}$, the difference $J(\underset{\sim}{\psi},\underset{\sim}{\varphi})$ does not generally vanish. Rather, we find from \eqref{J def} and \eqref{inhomogeneous} that
\begin{equation}\label{solvability basic}
{J}(\underset{\sim}{\psi},\underset{\sim}{\varphi})-\langle \underset{\sim}{f},\underset{\sim}{\varphi}\rangle=0.
\end{equation}

Using only the divergence theorem and that the flows associated with $\underset{\sim}{\psi}$ and $\underset{\sim}{\varphi}$ are incompressible, we have shown in \eqref{J general surface} that we can explicitly write ${J}(\underset{\sim}{\psi},\underset{\sim}{\varphi})$ in the form of surface integrals. Since ${J}(\underset{\sim}{\psi},\underset{\sim}{\varphi})$ vanishes for any $\underset{\sim}{\psi}$ in the natural domain of $\mathcal{L}$ and $\underset{\sim}{\varphi}$ in the natural domain of $\mathcal{L}^*$ (the adjoint property derived in Sec.~\ref{ssec:adjointoperator}), we expect that the apparent dependence in \eqref{solvability basic} upon the unknown pair $\underset{\sim}{\psi}$ can be removed. This would leave us with a necessary condition for solvability of \eqref{solvability basic} depending solely on the forcing terms, namely $\underset{\sim}{f}$ and any inhomogeneous terms in the auxiliary conditions satisfied by $\underset{\sim}{\psi}$. 

It will be useful to derive this condition for a class of inhomogeneous problems, formulated below,  which is more general than the inhomogeneous problem in Sec.~\ref{ssec:inhomogeneous}. The resulting solvability condition we shall find could then be applied to a wide range of scenarios, including but not limited to the perturbation scenarios we shall consider in Sec.~\ref{sec:perturbations}. 

To be consistent with the notation in Sec.~\ref{ssec:inhomogeneous}, we formulate the generalised inhomogeneous problem for the pair ${\underset{\sim}{\psi}}=(c_2\,\,\bu_2)^T$, with associated pressure $p_2$, stress tensor $\boldsymbol{\sigma}_2=\boldsymbol{S}[\bu_2]$ and particle velocities $\bU_2$ and $\boldsymbol{\Omega}_2$. We consider the forcing column vector $\underset{\sim}{f}= (\mathcal{C} \,\, \bzero)^T$, where $\mathcal{C}(\br)$ is a prescribed scalar field that does not grow too rapidly as $r\to\infty$ (in a manner to be specified later). This choice for $\underset{\sim}{f}$ means that the Stokes equations included in \eqref{inhomogeneous} are left homogeneous, a limitation which could easily be relaxed, if necessary. Next, we specify the form of the inhomogeneous auxiliary conditions satisfied by $\underset{\sim}{\psi}$. We consider the boundary conditions 
\refstepcounter{equation}
$$
\label{c bc forced gen}
\pd{c_2}{r}=\mathcal{A}, \quad \bu_2=\bnabla_s c_2+\boldsymbol{\Omega}_2\times\br + \boldsymbol{\mathcal{B}} \quad \text{at} \quad r=1
\eqno{(\theequation a,\!b)},
$$
where $\mathcal{A}$ and $\boldsymbol{\mathcal{B}}$ are scalar and vector functions of angular position, respectively, with the latter satisfying the impermeability constraint $\boldsymbol{\mathcal{B}}\bcdot\be_r=0$. 
The far-field condition on the concentration is prescribed by writing 
\begin{equation}\label{c far forced gen}
c_2=r\mathcal{R}+\mathscr{E},
\end{equation}
where $\mathcal{R}$ is a scalar function of angular position and the remainder $\mathscr{E}$ satisfies $\mathscr{E}=o(r)$ and $\bnabla\mathscr{E}=o(1)$ as $r\to\infty$. The far-field condition on the flow field is prescribed as
\begin{equation}\label{u far forced gen}
\bu_2\to-\bU_2 \quad \text{as} \quad r\to\infty.
\end{equation}
Lastly, we pose the integral constraints
\refstepcounter{equation}
$$
\label{F T balances forced gen}
\bF_2=\boldsymbol{\mathcal{F}}, \quad \bT_2=\boldsymbol{\mathcal{T}},
\eqno{(\theequation a,\!b)}
$$
in which $\boldsymbol{\mathcal{F}}$ and $\boldsymbol{\mathcal{T}}$ are arbitrary constant vectors. We note that the far-field condition \eqref{c far forced gen} is not specific enough to close the generalised inhomogeneous problem. As we shall see, it nevertheless suffices for the purpose of deriving a solvability condition. 

Consider the expression \eqref{J general surface} for ${J}(\underset{\sim}{\psi},\underset{\sim}{\varphi})$. Following the derivation of the adjoint property [cf.~\eqref{three lines} and \eqref{last line u}], we substitute the natural boundary conditions of $\mathcal{L}^*$ satisfied by $\underset{\sim}{\varphi}$ as well as the inhomogeneous auxiliary conditions \eqref{c bc forced gen}--\eqref{F T balances forced gen} satisfied by $\underset{\sim}{\psi}$. This leads to a simplified form for ${J}(\underset{\sim}{\psi},\underset{\sim}{\varphi})$, which we substitute into condition \eqref{solvability basic}. With $\underset{\sim}{\varphi}=(c'\,\,\bu')^T$ representing any of the adjoint homogeneous solutions \eqref{sol associated}, we find
\begin{multline}
\lim_{R\to\infty}\oint_{r=R}\mathrm{d}A\,\left(c'-r\pd{c'}{r}\right)\mathcal{R}-\oint_{r=1}\mathrm{d}A\, c'\mathcal{A}
\\+\oint_{r=1}\mathrm{d}A\,\boldsymbol{\mathcal{B}}\bcdot(\be_r\bcdot\boldsymbol{\sigma}')
-\bU'\bcdot\boldsymbol{\mathcal{F}}-\boldsymbol{\Omega}'\bcdot \boldsymbol{\mathcal{T}}=\lim_{R\to\infty}\int_{1<r<R}\mathrm{d}V\,c'\mathcal{C}.
\end{multline}
It remains to substitute the general form of the adjoint homogeneous solutions \eqref{sol associated} in order to make the above condition explicit. Note that the term involving the inhomogeneous torque $\boldsymbol{\mathcal{T}}$ vanishes trivially, since $\boldsymbol{\Omega}'=\bzero$, and that the term involving $\boldsymbol{\mathcal{B}}$ can be simplified using the result \eqref{stress result} upon recalling that $\boldsymbol{\mathcal{B}}$ has no radial component. Using the fact that $\bU'$ is arbitrary, we arrive at the explicit condition
\begin{equation}\label{solvability explicit}
3\lim_{R\to\infty}\frac{1}{R^2}\oint_{r=R}\mathrm{d}A\,\be_r\mathcal{R}-\oint_{r=1}\mathrm{d}A\, \be_r\mathcal{A}+ \oint_{r=1}\mathrm{d}A\,\boldsymbol{\mathcal{B}}
+\frac{1}{3}\boldsymbol{\mathcal{F}}=\lim_{R\to\infty}{\int_{1<r<R}\mathrm{d}V\,\be_r\frac{\mathcal{C}}{r^2}}.
\end{equation}
Note that the behavior of $\mathcal{C}$ as $r\to\infty$  must be such that the limit on the right-hand side of \eqref{solvability explicit} exists. 

Condition \eqref{solvability explicit} constitutes the main result of our adjoint method, which we shall apply in the next subsection to the canonical model and in the next section to a number of perturbation scenarios. We have here only shown that \eqref{solvability explicit} is a necessary condition for existence of a solution to the generalised inhomogeneous problem, by direct calculation. This represents one direction of the Fredholm Alternative theorem for differential operators \cite{Keener:18}, which if valid in the present setup would imply that \eqref{solvability basic}, and so \eqref{solvability explicit}, also constitute a sufficient condition for solvability of the generalised inhomogeneous problem. Rigorously proving the Fredholm Alternative in our setting, which involves an unbounded domain, as well as integral and incompressibility constraints, is outside the scope of our formal analysis. From a pragmatic point of view, a necessary condition suffices for the purpose of formally deriving nonlinear amplitude equations.

\subsection{Solvability condition for the canonical model}\label{ssec:basicsolvability}
We now apply the general solvability condition \eqref{solvability explicit} to the canonical model of an isotropic chemically active particle considered in the preceding section. Referring to the inhomogeneous problem of Sec.~\ref{ssec:inhomogeneous}, we have in that case that 
\refstepcounter{equation}
$$
\label{CR basic}
\mathcal{C}=4\bu_1\bcdot \bnabla c_1+\chi\bu_1\bcdot\bnabla\frac{1}{r}, \quad \mathcal{R}=2(\tI+\be_r\be_r)\boldsymbol{:}\bU_1\bU_1+4\be_r\bcdot\bU_1|\bU_1|,
\eqno{(\theequation a,\!b)}
$$
while $\mathcal{A}$, $\boldsymbol{\mathcal{B}}$ and $\boldsymbol{\mathcal{F}}$ vanish. 
With $c_1$ and $\bu_1$ given by \eqref{homogeneous sol}, we obtain 
\refstepcounter{equation}
$$
\label{CR integrals basic}
\lim_{R\to\infty}{\int_{1<r<R}\mathrm{d}V\,\be_r\frac{\mathcal{C}}{r^2}}=\chi \pi \bU_1, \quad \lim_{R\to\infty}\frac{1}{R^2}\oint_{r=R}\mathrm{d}A\,\be_r\mathcal{R}=\frac{16\pi}{3}\bU_1|\bU_1|.
\eqno{(\theequation a,\!b)}
$$
The amplitude equation \eqref{solvability isotropic} quoted in the preceding section readily follows from \eqref{solvability explicit}. 

\section{Perturbation scenarios}\label{sec:perturbations}
To further illustrate the adjoint method developed in the previous section, we go beyond the canonical model of an isotropic chemically active particle to consider the effects of perturbations about that model involving either the particle or its environment. The magnitude of each perturbations will be represented by some small positive parameter, which for the sake of discussion we here denote by $\delta$. For $\delta$ arbitrarily small, some perturbations have a leading-order effect sufficiently close to the instability threshold. In accordance with the notation used in Sec.~\ref{sec:isotropic}, we shall still use the small parameter $\epsilon$ defined by \eqref{chi def} to quantify the closeness of the P\'eclet number to the unperturbed critical value $\Pen=4$. Whereas in Sec.~\ref{sec:isotropic} we have carried out a local analysis in the limit $\Pen\to4$, i.e., $\epsilon\searrow0$, in the present section we shall analyze distinguished limits where both $\epsilon$ and $\delta$ are small, the smallness of $\epsilon$ relative to $\delta$ being such that the perturbation is just strong enough to influence the leading-order amplitude equations. In studying these distinguished limits, we shall slave $\delta$ to $\epsilon$ such that the parameter $\chi$ appearing in definition \eqref{chi def} will represent a rescaled and shifted bifurcation parameter that takes on arbitrary real values. This should be contrasted with the canonical isotropic scenario of Sec.~\ref{sec:isotropic}, where $\chi$ merely  indicated the sign of $\Pen-4$. 

\subsection{Uniform force and torque fields}
\subsubsection{Uniform force field}\label{sssec:force}
The first perturbation we consider is that of a uniform force field \cite{Yariv:17}. While this scenario has already been analysed in \cite{Saha:21}, our derivation here --- based on the adjoint  method --- has two advantages: (i) we do not need to solve the order-$\epsilon^2$ inhomogeneous problem, and (ii) we do not make the \textit{a priori} assumption that the particle motion is co-linear with the force field. The latter assumption, which is argued in \cite{Saha:21} on the basis of the bifurcation curves needing to match with intervals of $\Pen$ away from its critical value, follows here from a general three-dimensional analysis. As discussed in \cite{Saha:21}, this result corresponds to a steady-state alignment of the nominally isotropic spontaneous motion with the direction parallel or anti-parallel to the force field. 
\begin{figure}[t!]
\begin{center}
\includegraphics[scale=0.49]{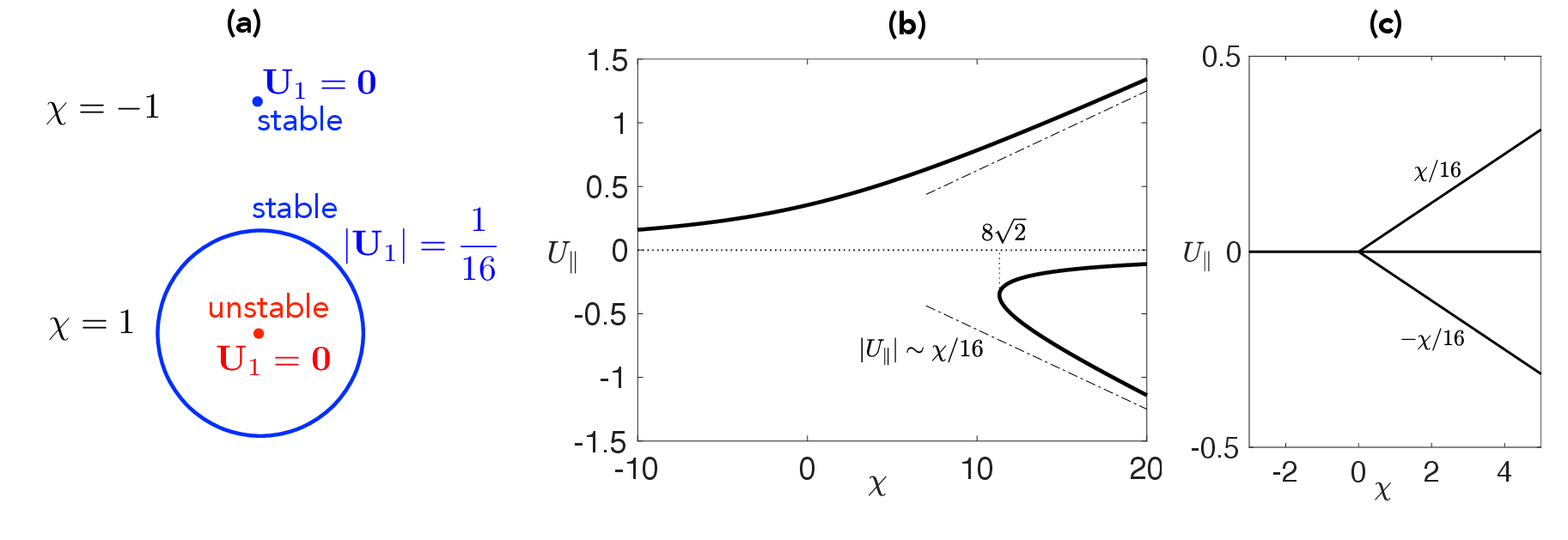}
\caption{Steady solutions: $\bU\sim \epsilon \bU_1$ as $\epsilon\searrow0$, with $\Pen=4+\epsilon\chi$ \textbf{(a)} \textbf{Canonical isotropic model} (Sec.~\ref{sec:isotropic}). $\bU_1$ satisfies the amplitude equation \eqref{solvability isotropic} describing a singular-pitchfork bifurcation from a symmetric-stationary state, which exists for all $\chi$, to spontaneous-motion states of magnitude $|\bU_1|=\chi/16$ and arbitrary direction, which exist for $\chi>0$. (In this case, it suffices to consider $\chi=\pm1$, without loss of generality.) Stability is indicated based on linear stability analysis of the stationary-symmetric state \cite{Michelin:13} and numerical simulations in \cite{Chen:21}. 
\textbf{(b)} \textbf{External force field} $6\pi\epsilon^2\unit$, where $\unit$ is a unit vector (Sec.~\ref{sssec:force}). Here $\bU_1=U_{\parallel}\unit$, with $U_{\parallel}$ satisfying the amplitude equation \eqref{force imperfect} describing an imperfect pitchfork bifurcation of spontaneous-motion states, which are restricted to the directions parallel or anti-parallel to the force field. \textbf{(c)} \textbf{External torque field} $8\pi\epsilon\unit$ (Sec.~\ref{sssec:torque}). We have $\bU_1=U_{\parallel}\unit$, with $U_{\parallel}$ satisfying the amplitude equation \eqref{torque perfect} describing a singular-pitchfork bifurcation of spontaneous-motion states restricted to the directions parallel or anti-parallel to the torque field; note that the spontaneous speed is identical to that in the isotropic canonical case.}
\label{fig:force}
\end{center}
\end{figure}

We shall represent the external force field by a dimensionless force $6\pi \bV$ acting on the particle, where $\bV$ is the dimensionless velocity that would be attained by the particle if it was chemically inert. With this convention, the force constraint (\ref{force torque balances}a)  becomes
\begin{equation}\label{force F}
\bF+6\pi\bV=\bzero.
\end{equation}
Let $\bV=V\unit$, where $V>0$ and $\unit$ is a unit vector, and consider the weak-force limit $V\ll1$ (in the present scenario, $V$ represents the small parameter $\delta$ discussed at the beginning of Sec.~\ref{sec:perturbations}). Inspecting the weakly nonlinear analysis in Sec.~\ref{sec:isotropic}, we see that the force first has a leading-order effect for $\epsilon=\mathcal{O}(\sqrt{V})$. We thus set $V=\epsilon^2$, without loss of generality. 

The only change to the weakly nonlinear analysis in Sec.~\ref{sec:isotropic} is in the order-$\epsilon^2$ inhomogeneous problem. The homogeneous force constraint (\ref{F T balances forced}a) becomes the inhomogeneous force constraint 
\begin{equation}\label{F2 force}
\bF_2=-6\pi\unit. 
\end{equation}
The general solvability condition \eqref{solvability explicit} applies as in the canonical scenario of Sec.~\ref{ssec:basicsolvability}, now with $\boldsymbol{\mathcal{F}}=-6\pi\unit$. We find the amplitude equation
\begin{equation}\label{bifurcation force field}
16\bU_1|\bU_1|-\chi\bU_1=2\unit.
\end{equation}

It is readily seen that the vector equation \eqref{bifurcation force field} reduces to the one-dimensional bifurcation relation obtained in \cite{Saha:21}. Indeed, it follows from \eqref{bifurcation force field} that $\bU_1={U}_{\parallel}\unit$, with ${U}_{\parallel}$ satisfying  
\begin{equation}\label{force imperfect}
16{U}_{\parallel}|{U}_{\parallel}|-\chi{U}_{\parallel}=2,
\end{equation}
in agreement with \cite{Saha:21}. The directionally restricted imperfect-pitchfork bifurcation relation implied by \eqref{force imperfect} is depicted in Fig.~\ref{fig:force}b. As  discussed in \cite{Saha:21}, the `parallel' branch (${U}_{\parallel}>0$), given by
\begin{equation}
{U}_{\parallel}=\frac{1}{32}\left(\chi+\sqrt{\chi^2+128}\right),
\end{equation} 
exists for all $\chi$. As $\chi\to-\infty$, this branch matches with solutions representing the linear response of the particle to the weak force field at P\'eclet numbers away from the threshold. As $\chi\to\infty$, it matches with the spontaneous-motion solutions in the canonical scenario when restricted to the direction parallel to the force field. For $\chi>8\sqrt{2}$, there are additionally two `anti-parallel' branches (${U}_{\parallel}<0$), given by
\begin{equation}
{U}_{\parallel}=\frac{1}{32}\left(-\chi\pm\sqrt{\chi^2-128}\right),
\end{equation} 
which are degenerate at $\chi=8\sqrt{2}$. As $\chi\to\infty$, one anti-parallel branch matches with solutions representing the linear response of the particle to the weak force field at P\'eclet numbers away from the threshold, while the other matches with the spontaneous-motion solutions in the canonical scenario when restricted to the direction anti-parallel to the force field. Stability of the above solution branches will be addressed in the subsequent part of this series. 

\subsubsection{Uniform torque field}\label{sssec:torque}
Consider now the effect of an external uniform torque field, represented by a dimensionless torque $8\pi\mathbf{W}$, in which $\mathbf{W}$ is the dimensionless angular velocity that would be attained by the particle if it was chemically inert. It is clear from the general solvability condition \eqref{solvability explicit} that an order-$\epsilon^2$ torque cannot influence the amplitude equation, in contrast to a force field at that order. An order-$\epsilon$ torque, however, would generate an order-$\epsilon$ rotational flow which could modify the order-$\epsilon^2$ inhomogeneous problem and thence the amplitude equation. To investigate this possibility, we set $\mathbf{W}=\epsilon \unit$, with $\unit$ a unit vector. Accordingly, the torque constraint (\ref{force torque balances}b) becomes
\begin{equation}\label{torque balance}
\mathbf{T}+8\pi \epsilon \unit=\bzero.
\end{equation}

The weakly nonlinear analysis of Sec.~\ref{sec:isotropic} is first modified at order $\epsilon$ of the particle-scale expansion. In light of \eqref{torque balance}, the homogeneous order-$\epsilon$ condition (\ref{force torque balances 1}b) is replaced by the inhomogeneous condition
\begin{equation}\label{T1 pT}
\mathbf{T}_1=-8\pi\unit.
\end{equation}
The general solution \eqref{homogeneous sol} to the order-$\epsilon$ particle-scale problem therefore needs to be supplemented by a particular solution that accounts for the right-hand side of \eqref{T1 pT}. A suitable particular solution is simply provided by the classical Stokes flow
\refstepcounter{equation}
$$
\label{rotational flow}
\bu_1\ub{p}=\unit\times \frac{\br}{r^3}, \quad p_1\ub{p}=0, 
\eqno{(\theequation a,\!b)}
$$
corresponding to the fluid velocity and pressure fields, respectively, that would be induced by the external torque if the sphere was chemically inert; this solution is associated with linear and angular particle velocities
\refstepcounter{equation}
$$
\label{rotational velocities}
\bU_1\ub{p} = \bzero,\quad \boldsymbol{\Omega}_1\ub{p} = \unit,
\eqno{(\theequation a,\!b)}
$$
respectively. 
Since the rotational flow field (\ref{rotational flow}a) vanishes in the radial direction, it cannot influence the coupled advection-diffusion equation \eqref{c1 eq}, so the particular solution for the concentration $c_1$ can be taken to vanish. Thus, the general solution to the order-$\epsilon$ problem now reads 
\refstepcounter{equation}
$$
\label{homogeneous sol pT}
c_1 = -2|\bU_1|+c_L(\br;\bU_1) , \quad \bu_{1}=\unit\times\frac{\br}{r^3}+\bu_L(\br;\bU_1), \quad p_{1}=p_L(\br;\bU_1), 
\eqno{(\theequation a\!\!-\!\!c)}
$$
with $c_L$, $\bu_L$ and $p_L$ provided by \eqref{linear sol}, $\bU_1$ arbitrary and 
\begin{equation}\label{Om1 pT}
\boldsymbol{\Omega}_1=\unit. 
\end{equation}

The order-$\epsilon^2$ particle-scale problem is the same as for the canonical isotropic model, only that $\bu_1$ is now provided by (\ref{homogeneous sol pT}b). In applying the general solvability condition \eqref{solvability explicit}, the modification to $\bu_1$ enters solely through the quantity $\mathcal{C}$, which is still defined by (\ref{CR basic}a). We find
\begin{equation}\label{Pdam C}
\lim_{R\to\infty}{\int_{1<r<R}\mathrm{d}V\,\be_r\frac{\mathcal{C}}{r^2}}= \chi \pi \bU_1 + \frac{22\pi}{5}\unit\times\bU_1,
\end{equation}
instead of (\ref{CR integrals basic}a). Otherwise, the calculation is the same as in Sec.~\ref{ssec:basicsolvability}, leading to the amplitude equation
\begin{equation}\label{torque amplitude}
16\bU_1|\bU_1|-\chi\bU_1=\frac{22}{5}\unit\times\bU_1.
\end{equation}

It follows from \eqref{torque amplitude} that $\bU_1=U_{\parallel}\unit$, where $U_{\parallel}$ satisfies
\begin{equation}\label{torque perfect}
U_{\parallel}(16|U_{\parallel}|-\chi) = 0, 
\end{equation}
implying the directionally restricted singular-pitchfork bifurcation depicted in Fig.~\ref{fig:force}c: we have the stationary-symmetric state, $U_{\parallel}=0$, for all $\chi$, as well as parallel and anti-parallel states, $U_{\parallel}=\pm 16/\chi$, for $\chi>0$. Thus, the external torque results in alignment, as in the force scenario, but without introducing imperfection, in contrast to the force scenario. 

\subsubsection{Parallel and perpendicular force and torque fields}\label{sssec:forcetorque}
Consider now the scenario where the particle is subjected to both an external force field $6\pi\bV$, of order $\epsilon^2$, and external torque field $8\pi\mathbf{W}$, of order $\epsilon$. Without loss of generality, we set $\bV=\epsilon^2\unit$, where $\unit$ is a prescribed unit vector, and $\mathbf{W}=\epsilon \mathbf{w}$, where $\mathbf{w}$ is a prescribed vector. Combining the analyses of the force and torque scenarios, we find 
\begin{equation}\label{crossed}
16\bU_1|\bU_1|-\chi\bU_1=2\unit+\frac{22}{5}\mathbf{w}\times\bU_1.
\end{equation}
In what follows we assume that neither the force nor torque vanish. In the case where the torque is parallel or anti-parallel to the force, it is readily seen that the torque has no effect, i.e., the steady states are the same as in the force scenario. We henceforth focus on the case where the force and torque fields are perpendicular. Let
\begin{equation}
\mathbf{w}=w\unitj,
\end{equation}
where $w>0$ and we introduce a right-handed orthogonal basis of unit vectors $(\unit,\unitj,\boldsymbol{\hat{k}})$. To simplify expressions, we define
\refstepcounter{equation}
$$
\label{crossed rescalings}
\bU_1=\frac{1}{\sqrt{8}}\boldsymbol{\mathcal{U}}, \quad \chi=\frac{16}{\sqrt{8}}\mathcal{X}, \quad w=\frac{10\sqrt{2}}{11}\mathcal{W},
\eqno{ (\theequation a\!\!-\!\!c)}
$$
whereby the amplitude equation \eqref{crossed} reads as
\begin{equation}\label{crossed reduced}
(\mathcal{U}-\mathcal{X})\boldsymbol{\mathcal{U}}=\unit+\mathcal{W}\unitj\times\boldsymbol{\mathcal{U}},
\end{equation}
with $\mathcal{U}=|\boldsymbol{\mathcal{U}}|$. 

\begin{figure}[t!]
\begin{center}
\includegraphics[trim={60 0 80 0}, scale=0.43]{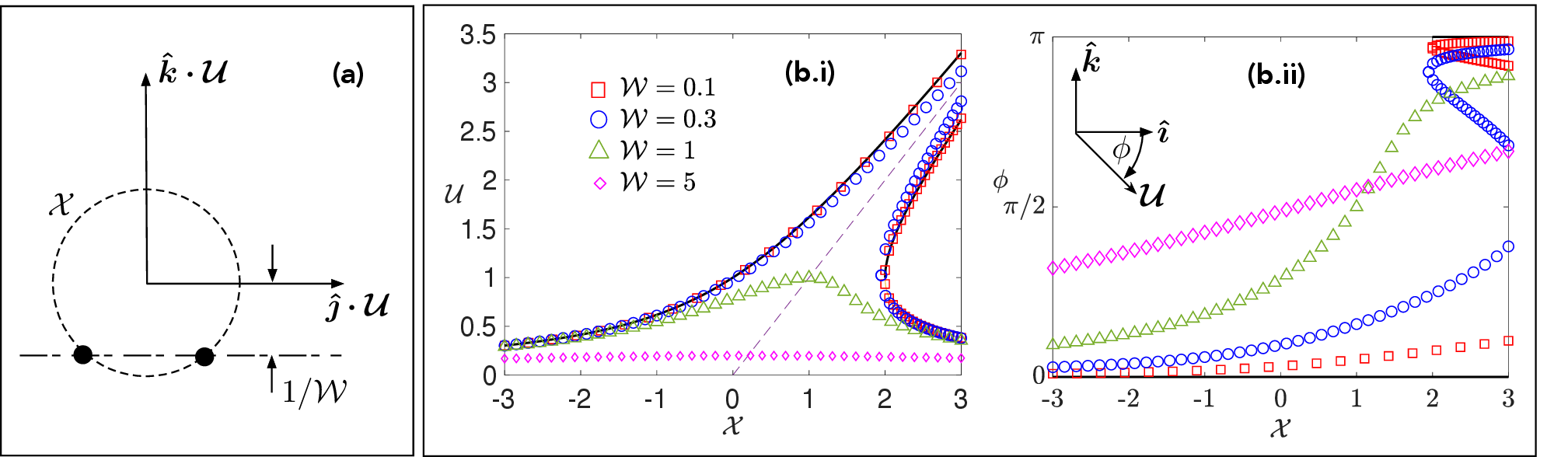}
\caption{Steady solutions $\bU\sim \epsilon\boldsymbol{\mathcal{U}}/\sqrt{8}$ as $\epsilon\searrow0$, with $\Pen=4+16\epsilon\mathcal{X}/\sqrt{8}$, in the scenario of a force field $6\pi\epsilon^2\unit$ perpendicular to a torque field $(80\pi\sqrt{2}/11) \epsilon \mathcal{W} \unitj$ (Sec.~\ref{sssec:forcetorque}). \textbf{(a)} There are up to two solutions perpendicular to the force, depending on the sign of $\mathcal{X}\mathcal{W}-1$. As $\mathcal{W}\nearrow\infty$, these solutions tend to the non-trivial states in the torque scenario. \textbf{(b)} There are between one and three solutions perpendicular to the torque. As $\mathcal{W}\searrow0$, these solutions limit to the solutions in the force scenario. As $\mathcal{W}\nearrow\infty$, there is one such solution that approaches the trivial state.}
\label{fig:crossed}
\end{center}
\end{figure}

We look for solutions $\boldsymbol{\mathcal{U}}$ of \eqref{crossed reduced} as a function of the real parameter $\mathcal{X}$ and the positive parameter $\mathcal{W}$. There are two families of solutions to consider. 
\begin{enumerate}
\item $\mathcal{U}=\mathcal{X}$. In that case, \eqref{crossed reduced} degenerates to 
\begin{equation}\label{simple}
\unit+\mathcal{W}\unitj\times\boldsymbol{\mathcal{U}}=\bzero,
\end{equation}
to be solved together with $\mathcal{U}=\mathcal{X}>0$. It is readily seen from \eqref{simple} that $\boldsymbol{\mathcal{U}}\perp \unit$, namely that the velocity is perpendicular to the force, and that $\boldsymbol{\hat{k}}\bcdot\boldsymbol{\mathcal{U}}=-1/\mathcal{W}$. As depicted schematically in Fig.~\ref{fig:crossed}a, the constraint $\mathcal{U}=\mathcal{X}$ then implies that there are zero solutions for $\mathcal{X}\mathcal{W}<1$, one for $\mathcal{X}\mathcal{W}=1$ and two for $\mathcal{X}\mathcal{W}>1$. Explicitly, those solutions are given by
\begin{equation}
\boldsymbol{\mathcal{U}}=\pm\left(\mathcal{X}^2-\frac{1}{\mathcal{W}^2}\right)^{1/2}\unitj-\frac{1}{\mathcal{W}}\boldsymbol{\hat{k}}.
\end{equation}
Note that $\boldsymbol{\mathcal{U}}\sim \pm \mathcal{X}\unitj$ as $\mathcal{W}\nearrow\infty$, i.e., these solutions limit to the non-trivial states in the torque scenario as the force magnitude vanishes relative to the torque magnitude. 

\item $\mathcal{U}\ne\mathcal{X}$. In that case, the $\unitj$ component of \eqref{crossed reduced} shows that $\boldsymbol{\mathcal{U}}\perp \unitj$, namely that the velocity is perpendicular to the torque. Inspecting the remaining components of \eqref{crossed reduced}, we find 
\begin{equation}\label{component form}
\boldsymbol{\mathcal{U}}=\mathcal{U}^2(\mathcal{U}-\mathcal{X})\unit-\mathcal{W}\mathcal{U}^2\boldsymbol{\hat{k}},
\end{equation}
where $\mathcal{U}>0$ satisfies the quartic equation 
\begin{equation}\label{U crossed eq}
\mathcal{U}^2\left[\left(\mathcal{U}-\mathcal{X}\right)^2+\mathcal{W}^2\right]-1=0.
\end{equation}
It is readily seen that \eqref{U crossed eq} always has at least one positive root, and that for $|\mathcal{X}|\le \sqrt{8}\mathcal{W}$ there cannot be additional positive roots. For $|\mathcal{X}|> \sqrt{8}\mathcal{W}$, \eqref{U crossed eq} has between one and three positive solutions. The existence of three positive roots for some choices of the parameters can be demonstrated by considering the limit $\mathcal{W}\searrow0$, where we find the solution
\begin{equation}\label{small W 1}
\mathcal{U}= \frac{1}{2}\left(\mathcal{X}+\sqrt{\mathcal{X}^2+4}\right)+\mathcal{O}\left(\mathcal{W}^2\right),
\end{equation}
which exists for arbitrary $\mathcal{X}$, as well as the pair of solutions 
\begin{equation}\label{small W 23}
\mathcal{U}= \frac{1}{2}\left(\mathcal{X}\pm\sqrt{\mathcal{X}^2-4}\right)+\mathcal{O}\left(\mathcal{W}^2\right),
\end{equation}
which, in the limit, exist for $\chi>2$ and are degenerate for $\chi=2$. In the opposite limit, $\mathcal{W}\nearrow\infty$, we find one positive root which vanishes in the limit like $\mathcal{U}\sim 1/\mathcal{W}$. In Fig.~\ref{fig:crossed}b.i, we plot the the positive roots of \eqref{U crossed eq} as a function of $\mathcal{W}$, for several values of $\mathcal{W}$, alongside the small-$\mathcal{W}$ approximations \eqref{small W 1} and \eqref{small W 23}. 

For any positive solution $\mathcal{U}$ of $\eqref{U crossed eq}$, the corresponding velocity vector $\boldsymbol{\mathcal{U}}$ is given by \eqref{component form}. It is more convenient to use the polar representation $\boldsymbol{\mathcal{U}}=\mathcal{U}(\unit\cos\phi-\boldsymbol{\hat{k}}\sin\phi)$, where  \eqref{component form} and \eqref{U crossed eq} together yield $\phi=\arccos\{\mathcal{U}(\mathcal{U}-\mathcal{X})\}\in(0,\pi)$. In Fig.~\ref{fig:crossed}b.ii, we plot $\phi$ corresponding to the $\mathcal{U}$ solutions shown in Fig.~\ref{fig:crossed}b.i. As $\mathcal{W}\searrow0$, $\phi\searrow 0$ for the solution \eqref{small W 1} and $\phi\nearrow\pi$ for the solutions \eqref{small W 23}.  As $\mathcal{W}\nearrow\infty$, the only solution satisfies $\phi\to\pi/2$.

Combining the small- and large-$\mathcal{W}$ limits of the magnitude $\mathcal{U}$ and angle $\phi$, we note the following regarding the torque-perpendicular solutions. As $\mathcal{W}\searrow0$, there are between one and three solutions that approach the states in the force scenario. As $\mathcal{W}\nearrow\infty$, the only solution approaches the trivial state. 
\end{enumerate}

To summarise, for perpendicular force and torque fields, there are between one and five solutions for the particle velocity vector. There are up to two that are perpendicular to the force, as depicted in Fig.~\ref{fig:crossed}a. These limit to the solutions in the torque scenario as the magnitude of the force is made negligible in comparison to that of the torque. Furthermore, there are between one and three that are perpendicular to the torque, as depicted in Fig.~\ref{fig:crossed}b. These limit to the solutions in the force scenario as the magnitude of the torque is made negligible in comparison to that of the force; in the opposite limit, there is one such solution that approaches the trivial state. 

\subsection{Non-uniform surface properties}
\label{ssec:properties}
\subsubsection{Steady formulation and amplitude equation}
We next consider weak, generally non-uniform, perturbations to the surface properties of the particle. Specifically, the dimensional prescribed flux and slip coefficient are modified as 
\refstepcounter{equation}
$$
\label{surface prop}
j_* \Rightarrow j_*(1-\epsilon^2\alpha), \quad b_*\Rightarrow b_*(1+\epsilon\beta),
\eqno{ (\theequation a,\!b)}
$$
where $\alpha$ and $\beta$ are functions of position on the particle boundary, fixed in surface coordinates attached to the particle; the flux and slip-coefficient perturbations have been scaled such that both have a leading-order effect as $\epsilon\searrow0$. In this part, we limit ourselves to solutions which appear steady in the co-moving frame introduced in Sec.~\ref{ssec:formu}, which does not rotate with the particle. In the present scenario, this implies that the particle does not rotate, or that its rotation is such that it leaves the surface properties fixed in the co-moving frame. There is also the possibility that the particle rotates sufficiently slowly such that the steady weakly nonlinear theory developed in this part holds in a quasi-static sense; this, however, requires an extremely small rotation rate, $\boldsymbol{\Omega}=o(\epsilon^2)$, to ensure that the associated time variation of $\bU_1$ is sufficiently slow such that the steady remote-region analysis in  Sec.~\ref{ssec:remote} remains valid in a quasi-static sense. In what follows, we shall apply our weakly nonlinear framework assuming a steady state in the co-moving frame, and then derive  conditions, associated with the rotation of the particle, for a given candidate solution to be consistent as a steady, or quasi-steady, solution. 

Given \eqref{surface prop}, we replace the boundary conditions \eqref{c bc} and \eqref{u bc} by
\refstepcounter{equation}
$$
\label{bcs Pproperties}
\pd{c}{r}=-1+\epsilon^2 \alpha, \quad \bu=(1+\epsilon\beta)\bnabla_s c +\mathbf{\Omega}\times \br \quad \text{at} \quad r=1,
\eqno{ (\theequation a,\!b)}
$$
respectively. The only change to the weakly nonlinear analysis in Sec.~\ref{sec:isotropic} is in the order-$\epsilon^2$ inhomogeneous problem, where the boundary conditions \eqref{c bc forced} and \eqref{u bc forced} are replaced by
\refstepcounter{equation}
$$
\label{bcs properties}
\pd{c_2}{r}=\alpha, \quad \bu_2=\bnabla_sc_2+\boldsymbol{\Omega}_2\times\br + \beta\bnabla_s c_1
 \quad \text{at} \quad r=1,
 \eqno{(\theequation a,\!b)}
$$
respectively. Applying the general solvability condition \eqref{solvability explicit}, with 
\refstepcounter{equation}
$$
\label{AB properties}
\mathcal{A}=\alpha, \quad \boldsymbol{\mathcal{B}}=\beta\bnabla_s c_1,
 \eqno{(\theequation a,\!b)}
$$ 
and $\mathcal{R}$ and $\mathcal{C}$ as in Sec.~\ref{ssec:basicsolvability}, we find the amplitude equation
\begin{equation}\label{bifurcation properties}
16\bU_1|\bU_1|-\chi\bU_1-\frac{3}{2\pi}\left\{\oint_{r=1}\mathrm{d}A\,(\tI-\be_r\be_r)\beta\right\}\bcdot\bU_1 = \frac{1}{\pi}\oint_{r=1}\mathrm{d}A\,\be_r\alpha,
\end{equation}
where we have used (\ref{AB properties}b), with (\ref{homogeneous sol}a), to show that 
\begin{equation}
\oint_{r=1}\mathrm{d}A\,\boldsymbol{\mathcal{B}}=\left\{-\frac{3}{2}\oint_{r=1}\mathrm{d}A\,(\tI-\be_r\be_r)\beta\right\}\bcdot\bU_1.
\end{equation}

We see from  \eqref{bifurcation properties} that the effect of the flux perturbation is similar to that of an external force field, the equivalent force field being proportional to a dipole moment of the flux perturbation. The slip-coefficient perturbation modifies the homogeneous linear term, effectively adding a general second-order tensor to the bifurcation parameter $\chi$.

To check whether a solution $\bU_1$ to the amplitude equation \eqref{bifurcation properties} is consistent, we must consider the particle's angular rotation. Since $\boldsymbol{\Omega}_1$ vanishes trivially in the present scenario, and since $\boldsymbol{\Omega}=o(\epsilon^2)$ is permissible, we need only consider $\boldsymbol{\Omega}_2$. Fortunately, it is possible to calculate $\boldsymbol{\Omega}_2$ without solving the generalised order-$\epsilon^2$ inhomogeneous problem in detail. Following  \cite{Stone:96}, we apply the Lorenz Reciprocal Theorem to the Stokes problem included in the order-$\epsilon^2$ inhomogeneous problem, obtaining $\boldsymbol{\Omega}_2$ as a functional of the relative fluid velocity at the surface (cf.~(\ref{bcs properties}b)): 
\begin{equation}
\boldsymbol{\Omega}_2=-\frac{3}{8\pi}\oint_{r=1}\mathrm{d}A\,\be_r\times \left(\bnabla_sc_2+\boldsymbol{\mathcal{B}}\right).
\end{equation}
While we do not know $c_2$, the associated contribution can be shown to vanish trivially using a Stokes-type integral theorem (see Appendix \ref{app:surfaceop}; the physical fact that slip uniformly proportional to the surface gradient of a scalar field cannot drive particle rotation was pointed out to me by Ehud Yariv). Accordingly, substitution of (\ref{AB properties}b), with (\ref{homogeneous sol}a), yields 
\begin{equation}\label{om 2 prop}
\boldsymbol{\Omega}_2=-\frac{9}{16\pi}\bU_1\times\oint_{r=1} \mathrm{d}A\,\beta\be_r.
\end{equation}
A solution $\bU_1$ is consistent only if $\boldsymbol{\Omega}_2$ vanishes or is such that it leaves the surface distributions $\alpha$ and $\beta$ fixed in the co-moving frame. Otherwise, the first term in the amplitude equation \eqref{bifurcation properties} is wrong, as it originates from matching with an inconsistent  steady solution in the remote region. 

\subsubsection{Axisymmetric perturbation}
\label{sssec:axiprop}
For the sake of illustration, we henceforth focus on surface perturbations that are symmetric about an axis that points in the direction of the unit vector $\hat{\bp}$ and passes through the particle's centroid, such that $\alpha$ and $\beta$ are functions of $\cos\theta=\hat{\bp}\bcdot\be_r$, with $0\le\theta\le\pi$. We begin by considering the angular rotation in this case. As a consequence of the axial symmetry, we can write
\begin{equation}
\frac{9}{16\pi}\oint_{r=1}\mathrm{d}A\,\be_r\beta=\beta_r\hat{\bp},
\end{equation}
where we define
\begin{equation}\label{beta r def}
\beta_r=\frac{9}{8}\int_0^\pi\mathrm{d}\theta\,\beta(\theta)\sin\theta\cos\theta.
\end{equation}
We then find from \eqref{om 2 prop} that 
\begin{equation}
\boldsymbol{\Omega}_2=\beta_r \hat{\bp}\times\bU_1. 
\end{equation}
The fact that $\boldsymbol{\Omega}_2$ is perpendicular to $\hat{\bp}$ rules out the possibility of rotation that leaves the axisymmetric surface distributions fixed in the co-moving frame. A consistent steady (or quasi-steady) state therefore requires $\boldsymbol{\Omega}_2=\bzero$, implying that non-longitudinal solutions ($\bU_1\times\hat{\bp}\ne\bzero$) only represent consistent steady states if $\beta_r=0$. We note that $\beta_r$ vanishes trivially for fore-aft symmetric slip-coefficient perturbations, i.e., when $\beta(\theta)=\beta(\pi-\theta)$. 

\begin{figure}[t!]
\begin{center}
\includegraphics[scale=0.35]{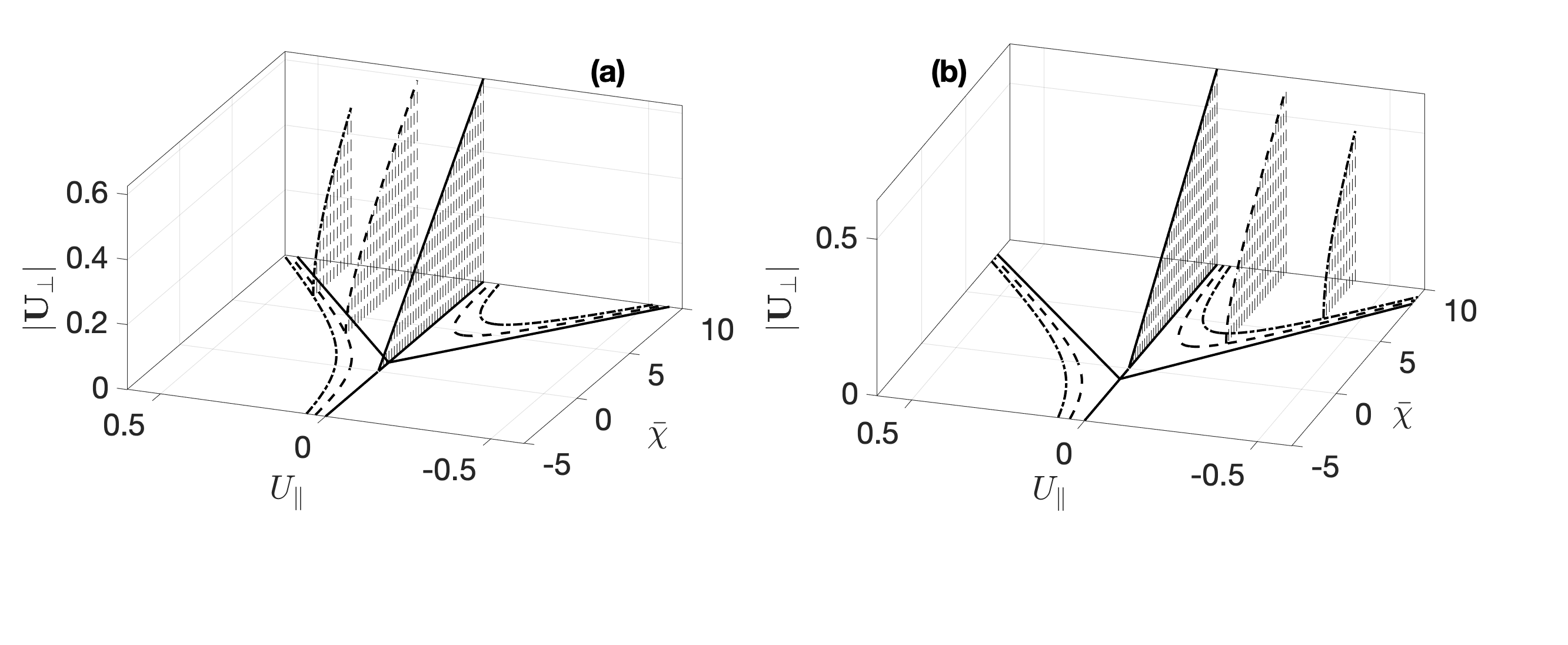}
\caption{Steady solutions $\bU\sim\epsilon(U_{\parallel}\hat{\bp}+\bU_{\perp})$ as $\epsilon\searrow0$, with $\Pen=4+\epsilon\chi$ and $\bU_{\perp}\bcdot\hat{\bp}=0$, in the scenario of flux and slip-coefficient perturbations symmetric about an axis defined by the unit vector $\hat{\bp}$ and the particle's centroid (Sec.~\ref{sssec:axiprop}). Non-longitudinal solutions with $\bU_{\perp}\ne\bzero$ are only consistent in the case $\beta_r=0$ (cf.~\eqref{beta r def}); for such solutions, the direction of $\bU_{\perp}$ normal to $\hat{\bp}$ is arbitrary. Solution branches are shown as a function of the shifted bifurcation parameter $\bar{\chi}=\chi+\beta_I$, for $\bar{\alpha}=0$ (solid lines), $\bar{\alpha}=0.2$ (dashed curves) and $\bar{\alpha}=0.4$ (dash-dotted curves), in the two cases (\textbf{a}): $\bar{\beta}=-1$, and (\textbf{b}): $\bar{\beta}=1$. For definitions of the lumped perturbation parameters $\bar{\alpha},\bar{\chi}$ and $\bar{\beta}$, see \eqref{bar alpha def} and \eqref{betas def}.} 
\label{fig:properties}
\end{center}
\end{figure}
Consider now how the amplitude equation \eqref{bifurcation properties} simplifies for axisymmetric perturbations. Noting that, as a consequence of the axial symmetry, we have 
\refstepcounter{equation}
$$
\label{bars def}
\frac{1}{\pi}\oint_{r=1}\mathrm{d}A\,\be_r\alpha= \bar{\alpha}\hat{\bp}, \quad \frac{3}{2\pi}\oint_{r=1}\mathrm{d}A\,(\tI-\be_r\be_r)\beta=\beta_I\tI +\bar{\beta}\hat{\bp}\hat{\bp},
\eqno{(\theequation a,\!b)}
$$
where we define
\begin{equation}\label{bar alpha def}
\bar{\alpha}=2\int_0^\pi\mathrm{d}\theta\,\alpha(\theta)\sin\theta\cos\theta
\end{equation}
and 
\refstepcounter{equation}
$$
\label{betas def}
\beta_I=\frac{3}{4}\int_0^{\pi}\mathrm{d}\theta\,\beta(\theta)\sin\theta(3+\cos2\theta), \quad \bar{\beta}=-\frac{3}{4}\int_0^{\pi}\mathrm{d}\theta\,\beta(\theta)\sin\theta(1+3\cos2\theta),
 \eqno{(\theequation a,\!b)}
$$
we find the amplitude equation 
\begin{equation}
16\bU_1|\bU_1|-\bar{\chi}\bU_1-\bar{\beta}\hat{\bp}\hat{\bp}\bcdot\bU_1=\bar{\alpha}\hat{\bp},
\end{equation}
with $\bar{\chi}=\chi+\beta_I$ being a shifted bifurcation parameter, that subsumes the isotropic effect of the parameter $\beta_I$. 

The case $\bar{\beta}=0$ is analogous to the force scenario (cf.~\eqref{bifurcation force field}). In particular, since that scenario involves only longitudinal solutions, we need not worry about particle rotation rendering those solutions inconsistent. 

Consider next the case $\bar{\beta}\ne0$, where without loss of generality we assume $\bar{\alpha}\ge0$ and $\bar{\beta}=\pm1$. To identify all solutions branches, we write $\bU_1=U_{\parallel}\hat{\bp}+\bU_{\perp}$, where $\bU_{\perp}\bcdot\hat{\bp}=0$, and consider below the sub-cases $\bar{\alpha}=0$ and $\bar{\alpha}>0$. 
\begin{enumerate}
\item For $\bar{\alpha}=0$, we find the following solution branches:
\begin{enumerate}
\item For all $\bar{\chi}$, there is the trivial solution $\bU_1=\bzero$, which corresponds to the stationary-symmetric state. 
\item For $\bar{\chi}>-\bar{\beta}$, there are two longitudinal solutions ($\bU_{\perp}=\bzero$), with
\begin{equation}
U_{\parallel}=\pm\frac{\bar{\chi}+\bar{\beta}}{16}.
\end{equation}
\item For $\bar{\chi}>0$, there are transverse solutions ($U_{\parallel}=0$), with $\bU_{\perp}$ in an arbitrary direction perpendicular to $\hat{\bp}$ and magnitude
\begin{equation}
|\bU_{\perp}|=\frac{\bar{\chi}}{16}.  
\end{equation}
Note that the longitudinal and transverse branches both bifurcate from the stationary-symmetric state; the transverse bifurcation precedes the longitudinal bifurcation for $\bar{\beta}=-1$ and follows it for $\bar{\beta}=1$. These transverse solutions are only consistent in the case $\beta_r=0$. 
\end{enumerate}
\item For $\bar{\alpha}>0$, we find the following solution branches:
\begin{enumerate}
\item Longitudinal solutions ($\bU_{\perp}=\bzero$), which satisfy the one-dimensional bifurcation relation
\begin{equation}
\left(16|U_{\parallel}|-\bar{\chi}-\bar{\beta}\right)U_{\parallel}=\bar{\alpha},
\end{equation}
with one parallel ($U_{\parallel}>0$) branch
\begin{equation}
U_{\parallel}=\frac{1}{32}\left(\bar{\chi}+\bar{\beta}+\sqrt{(\bar{\chi}+\bar{\beta})^2+64\bar{\alpha}}\right),
\end{equation}
which exists for all $\bar{\chi}$, and two anti-parallel ($U_{\parallel}<0$) branches
\begin{equation}
U_{\parallel}=-\frac{1}{32}\left(\bar{\chi}+\bar{\beta}\pm\sqrt{(\bar{\chi}+\bar{\beta})^2-64\bar{\alpha}}\right). 
\end{equation}
which exist for $\bar{\chi}+\bar{\beta}\ge8\bar{\alpha}^{1/2}$, these solutions being  degenerate for the equality.

\item Transverse-longitudinal solutions of magnitude $|\bU_1|=\bar{\chi}/16$, with longitudinal and transverse parts 
\refstepcounter{equation}
$$
U_{\parallel}=-\frac{\bar{\alpha}}{\bar{\beta}}, \quad |\bU_{\perp}|=\left(\frac{\chi}{16}+\bar{\alpha}\right)^{1/2}\left(\frac{\chi}{16}-\bar{\alpha}\right)^{1/2},
\eqno{(\theequation a,\!b)}
$$
the direction of $\bU_{\perp}$ being normal to $\hat{\bp}$ but otherwise arbitrary. These solutions exist for $\bar{\chi}>16\bar{\alpha}$. The mixed transverse-longitudinal solutions become longitudinal as $\bar{\chi}\searrow 16\bar{\alpha}$; for $\bar{\beta}=-1$, they bifurcate from the parallel solution branch, while for $\bar{\beta}=1$ they bifurcate from one of the anti-parallel solution branches. These transverse-longitudinal  solutions are only consistent in the case $\beta_r=0$. 
\end{enumerate}
\end{enumerate}
In Fig.~\ref{fig:properties}, we depict the above solution branches obtained for $\bar{\beta}\ne0$, for both $\bar{\beta}=\pm1$ and for several values of $\bar{\alpha}$. In the subsequent part, we will investigate the stability of the above solution branches, and explore the dynamical ramifications of having non-longitudinal solutions that cease to be consistent steady states for $\beta_r\ne0$. 

\subsection{First-order surface kinetics}
In the following two subsections we consider perturbations to the chemical model of the particle and the liquid solution, respectively. Unlike the first two perturbations considered in this section, these perturbations retain the isotropy of the canonical model of  Sec.~\ref{sec:isotropic}. 

In this subsection, we consider a generalised model for the chemical activity at the surface of the particle. We assume that, in addition to the constant and uniform supply of solute molecules at the particle boundary, represented by the positive flux $j_*$, solute molecules are also absorbed at the surface according to a first-order chemical reaction \cite{Michelin:14}. The total surface flux can be written as
$$
\text{solute flux}=j_*-k_*\times (\text{concentration at surface}),
$$
where $k_*$ is a rate constant. With $\bar{c}_*$ the concentration at infinity as in Sec.~\ref{ssec:formu}, we assume that $j'_*=j_*-k_*\bar{c}_*$ is positive and normalise the concentration deviation from $\bar{c}_*$ by $c'_*=a_*j'_*/D_*$, which modifies the characteristic concentration $c_*$ defined in Sec.~\ref{ssec:formu}. The dimensionless problem is then identical to that formulated in Sec.~\ref{ssec:formu}, except that the boundary condition \eqref{c bc} is replaced by
\begin{equation}\label{c bc Pdam}
\pd{c}{r}=-1+\mathrm{Da}_s c \quad \text{at} \quad r=1,
\end{equation}
with $\mathrm{Da}_s=k_*a_*/D_*$ a `surface' Damkohler number (distinguished from the `bulk' Damkohler we shall introduce in the following subsection). Comparing with \eqref{c bc}, the prescribed-flux assumption associated with the canonical model corresponds to the limit $\mathrm{Da}_s\searrow0$. We shall demonstrate that for arbitrarily small $\mathrm{Da}_s$, the surface absorption is important for $\Pen$ sufficiently close to its critical value. 

Naively, the form of \eqref{c bc Pdam} suggests the distinguished interval $\epsilon=\mathcal{O}(\sqrt{\mathrm{Da}_s})$, since then the new term $\mathrm{Da}_s c$ enters the order-$\epsilon^2$  inhomogeneous problem. Owing to $c_0$ being isotropic, however, such modification of the inhomogeneous problem would not have any effect on the solvability condition. Our analysis below will confirm that the relevant distinguished interval is, in fact, $\epsilon=\mathcal{O}(\mathrm{Da}_s)$. We accordingly set $\mathrm{Da}_s=\epsilon$, without loss of generality.  

In the above regime, the weakly nonlinear expansion is modified already at order $\epsilon$. Recalling that $c_0=1/r$, it follows from \eqref{c bc Pdam} that the homogeneous boundary condition \eqref{c1 bc} is replaced by the inhomogeneous boundary condition 
\begin{equation}\label{c1 bc Pdam}
\pd{c_1}{r}=1 \quad \text{at} \quad r=1.
\end{equation}
Otherwise, the order-$\epsilon$ problem is the same as in the canonical scenario. Its general solution can therefore be obtained by adding to the general homogeneous solution \eqref{homogeneous sol} a particular solution accounting for the new right-hand side in \eqref{c1 bc Pdam}. It is easy to see that a suitable particular solution consists of an isotropic $-1/r$ concentration field that does not generate flow. The general solution to the order-$\epsilon$ problem is therefore
\refstepcounter{equation}
$$
\label{homogeneous sol Psurf}
c_1 = -\frac{1}{r}-2|\bU_1|+c_L(\br;\bU_1) , \quad \bu_{1}=\bu_L(\br;\bU_1), \quad p_{1}=p_L(\br;\bU_1), 
\eqno{(\theequation a\!\!-\!\!c)}
$$
instead of (\ref{homogeneous sol}).

We are now ready to consider the inhomogeneous problem at order $\epsilon^2$ of the weakly nonlinear expansion. There are two changes relative to the problem formulated in Sec.~\ref{ssec:inhomogeneous} for the canonical model. First, it follows from \eqref{c bc Pdam} that the boundary condition \eqref{c bc forced} is replaced by
\begin{equation}\label{c2 bc Pdam}
\pd{c_2}{r}=c_1 \quad \text{at} \quad r=1.
\end{equation}
The second is that $c_1$ in \eqref{c2 bc Pdam} as well as in the concentration equation \eqref{c eq forced} are given by (\ref{homogeneous sol Psurf}a) rather than (\ref{homogeneous sol}a).

It remains to apply the solvability condition \eqref{solvability explicit} to the modified inhomogeneous problem at order $\epsilon^2$. There are two changes relative to the canonical case considered in Sec.~\ref{ssec:basicsolvability}. First, the quantity $\mathcal{C}$ retains the form (\ref{CR basic}a) but with $c_1$ now given by (\ref{homogeneous sol Psurf}a). With the change to $\mathcal{C}$ being equivalent to subtracting $4$ from $\chi$, (\ref{CR integrals basic}a) is replaced by 
\begin{equation}\label{Pdam C}
\lim_{R\to\infty}{\int_{1<r<R}\mathrm{d}V\,\be_r\frac{\mathcal{C}}{r^2}}= (\chi-4) \pi \bU_1.
\end{equation}
Second, the boundary condition \eqref{c2 bc Pdam} gives $\mathcal{A}=c_1$, in which $c_1$ is given by (\ref{homogeneous sol Psurf}a) evaluated at $r=1$. We find
\begin{equation}\label{Pdam A}
\oint_{r=1}\mathrm{d}A\, \be_r\mathcal{A}=-2\pi\bU_1.
\end{equation}

With \eqref{Pdam C} and \eqref{Pdam A}, the solvability condition \eqref{solvability explicit} yields the amplitude equation
\begin{equation}\label{bifurcation Pdam}
\bU_1\left(16|\bU_1|-\chi+6\right)=0.
\end{equation}
Comparing with \eqref{solvability isotropic}, the bifurcation  is similar to that in the canonical case except for a shift to a higher P\'eclet number. In unscaled notation, \eqref{bifurcation Pdam} implies that for $\Pen>4+6\mathrm{Da}_s+o(\mathrm{Da}_s)$ there is steady spontaneous motion in an arbitrary direction with magnitude having the local behavior
\begin{equation}\label{bifurcation unscaled Pdam}
|\bU|\sim \frac{\Pen-4-6\mathrm{Da}_s}{16} \quad \text{as} \quad \mathrm{Da}_s\searrow0, \quad \text{with} \quad \Pen-4=\mathcal{O}(\mathrm{Da}_s).
\end{equation}

\subsection{Bulk absorption}\label{ssec:bulk}
As a final example we consider the effect of solute absorption in the liquid bulk  \cite{De:13}. We assume that the solute is absorbed in proportion to the deviation of the concentration from the equilibrium value $\bar{c}_*$. We denote the absorption rate by $\kappa_*$ and adopt the same dimensionless notation as in Sec.~\ref{sec:isotropic}. The dimensionless formulation of the problem is then the same as in Sec.~\ref{ssec:formu} except that the advection-diffusion equation \eqref{c eq} becomes \cite{De:13}
\begin{equation}\label{c eq Pbulk}
\nabla^2c-\Pen \bu\bcdot \bnabla c = {\mathrm{Da}_b}c,
\end{equation}
where we define a `bulk' Damkohler number $\mathrm{Da}_b=\kappa_*a_*^2/D_*$. Naively, the leading-order absorption $\sim -\mathrm{Da}_b c_0$ in the particle region suggests that bulk reactions enter the inhomogeneous order-$\epsilon^2$ problem for $\epsilon=\mathcal{O}(\sqrt{\mathrm{Da}_b})$. While that is indeed the appropriate distinguished scaling, the isotropic absorption associated with $c_0$ does not actually influence the solvability condition. We shall see that it is rather  the effect of bulk reactions in the remote region that, for $\epsilon=\mathcal{O}(\sqrt{\mathrm{Da}_b})$, influence the solvability condition via a modified far-field condition in the order-$\epsilon^2$ inhomogeneous problem. Without loss of generality, we set ${\mathrm{Da}}_b=\epsilon^2$. 

We must first consider how the analysis in Sec.~\ref{ssec:remote} of the remote region is modified by bulk reactions. Adopting the same definitions as in Sec.~\ref{ssec:remote}, we find that the  the leading-order advection-diffusion equation \eqref{remote eq} becomes the advection-diffusion-reaction equation 
\begin{equation}\label{remote eq Pbulk}
\tilde{\nabla}^2\tilde{c}_1+4\bU_1\bcdot\tilde{\bnabla}\tilde{c}_1=\tilde{c}_1,
\end{equation}
to be solved together with the decay condition \eqref{remote far} and the matching condition \eqref{remote matching}, as in the canonical isotropic scenario. The latter condition is not modified since it is relies on matching with the leading-order particle-scale concentration $c_0$, which is not affected by the weak bulk absorption. The solution to the modified remote problem is readily found as 
\begin{equation}\label{remote sol Pbulk}
\tilde{c}_1 = \frac{1}{\tilde{r}}\exp\left\{-2\bU_1\bcdot\tilde{\br}-\tilde{r}\sqrt{4|\bU_1|^2+1}\right\}, 
\end{equation}
which replaces \eqref{remote sol}. 

The modified leading-order solution in the remote region implies, through asymptotic matching, modified far-field conditions at orders $\epsilon$ and $\epsilon^2$ of the particle-region expansion. At order $\epsilon$ we find that \eqref{c1 far} is generalised as 
\begin{equation}\label{c1 far Pbulk}
c_1 = -2\be_r\bcdot\bU_1-\sqrt{4|\bU_1|^2+1} + o(1) \quad \text{as} \quad r\to\infty,
\end{equation}
while at order $\epsilon^2$ we find that \eqref{c far forced} is generalised as
\begin{equation}\label{c2 far Pbulk}
c_2 = r\left\{2(\tI+\be_r\be_r)\boldsymbol{:}\bU_1\bU_1+2\be_r\bcdot \bU_1 \sqrt{4|\bU_1|^2+1}+\frac{1}{2}\right\}+ o(r) \quad \text{as} \quad r\to\infty.
\end{equation}

As a consequence of \eqref{c1 far Pbulk}, the general solution to the order-$\epsilon$ particle-region problem is modified from (\ref{homogeneous sol}) to 
\refstepcounter{equation}
$$
c_1 = -\sqrt{4|\bU_1|^2+1}+c_L(\br;\bU_1) , \quad \bu_{1}=\bu_L(\br;\bU_1), \quad p_{1}=p_L(\br;\bU_1),
\eqno{(\theequation a\!\!-\!\!c)}
$$ 
The only difference is in the reference value of the concentration $c_1$; we will see that this has no effect. 

Turning to the inhomogeneous problem at order $\epsilon^2$, the inhomogeneous coupled advection-diffusion equation \eqref{c eq forced} becomes
\begin{equation}\label{c eq forced Pb}
\nabla^2c_2 -4\bu_2\bcdot\bnabla\frac{1}{r}= 4\bu_1\bcdot \bnabla c_1 + \chi\bu_1\bcdot\bnabla\frac{1}{r}+\frac{1}{r}.
\end{equation}
The quantity $\mathcal{C}$ appearing in the solvability condition \eqref{solvability explicit} is defined as the right-hand side of \eqref{c eq forced Pb} --- this is just (\ref{CR basic}a) plus the isotropic contribution $1/r$, which is readily seen to have no effect on the solvability condition. In contrast, the far-field condition \eqref{c2 far Pbulk} modifies $\mathcal{R}$ from (\ref{CR basic}b) to the expression in the curly brackets of \eqref{c2 far Pbulk}; we accordingly find that (\ref{CR integrals basic}b) generalises as
\begin{equation}
\lim_{R\to\infty}\frac{1}{R^2}\oint_{r=R}\mathrm{d}A\,\be_r\mathcal{R}=\frac{8\pi}{3}\bU_1\sqrt{4|\bU_1|^2+1}.
\end{equation}
\begin{figure}[t!]
\begin{center}
\includegraphics[scale=0.45]{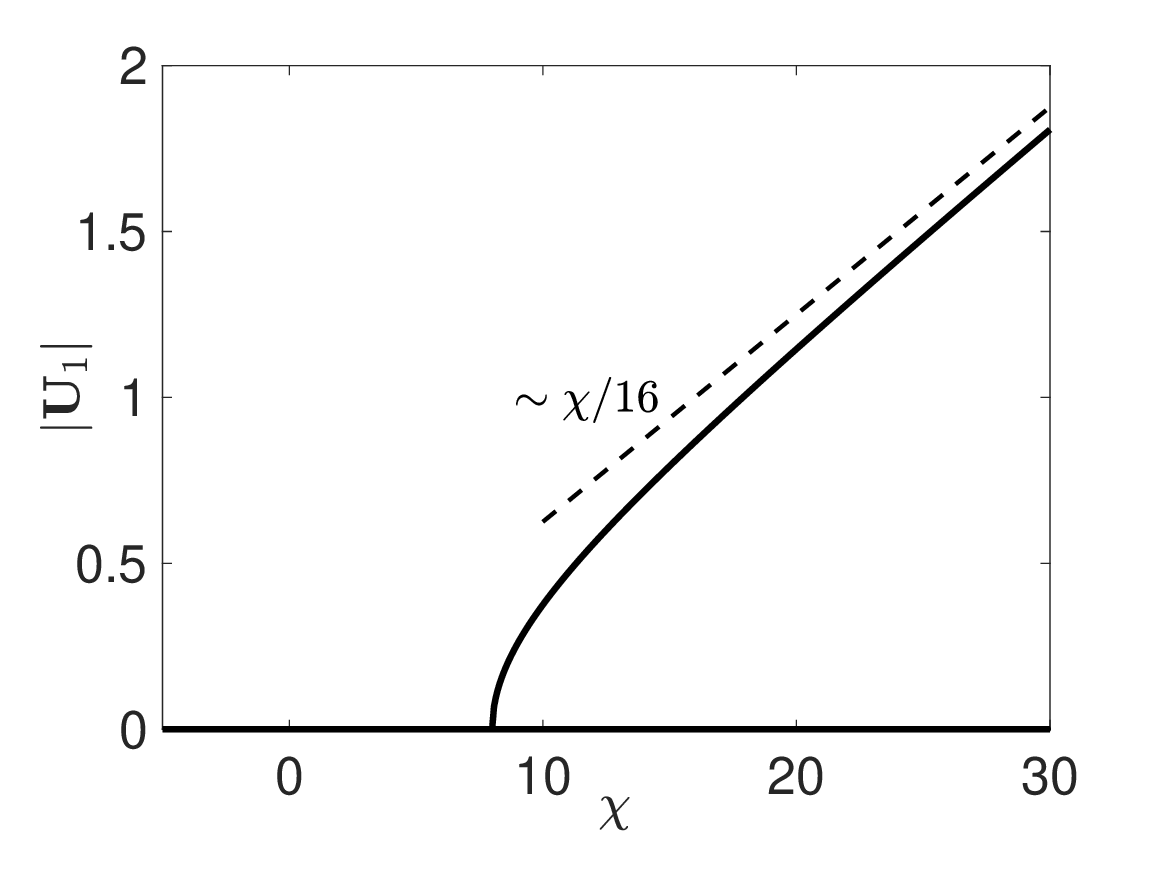}
\caption{Effect of bulk absorption with Damkohler number $\mathrm{Da}_b\ll1$ (Sec.~\ref{ssec:bulk}). Steady solutions $\bU\sim \sqrt{\mathrm{Da}_b}\mathbf{U}_1$, with $\Pen=4+\chi\sqrt{\mathrm{Da}_b}$, $\bU_1$ having an arbitrary direction and magnitude as plotted as a function of $\chi$; as $\chi\to\infty$, the solutions approach those in the canonical isotropic model.}
\label{fig:bulk}
\end{center}
\end{figure}

Substituting the above results into the solvability condition \eqref{solvability explicit}, we find the amplitude equation
\begin{equation}\label{bifurcation Pbulk}
\bU_1\left(16\sqrt{|\bU_1|^2+1/4}-\chi\right)=\bzero,
\end{equation}
which generalises \eqref{solvability isotropic}. The trivial solution $\bU_1=\bzero$ exists for all $\chi$. For $\chi>8$ there are also spontaneous-motion states having magnitude 
\begin{equation}
|\bU_1|=\frac{1}{16}\sqrt{\chi^2-64}
\end{equation}
and arbitrary direction. In unscaled notation, \eqref{bifurcation Pbulk} implies that for $\Pen>4+8\sqrt{\mathrm{Da}_b}+o(\sqrt{\mathrm{Da}_b})$ the particle can sustain steady rectilinear motion in an arbitrary direction with the speed having the local behavior
\begin{equation}\label{bifurcation unscaled Pbulk}
|\bU|\sim \frac{1}{16}\sqrt{(\Pen-4)^2-64\mathrm{Da}_b} \quad \text{as} \quad \mathrm{Da}_b\searrow0, \quad \text{with} \quad \Pen-4=\mathcal{O}\left(\sqrt{\mathrm{Da}_b}\right).
\end{equation}

The bifurcation relation \eqref{bifurcation Pbulk} is depicted in Fig.~\ref{fig:bulk}. As already noted in \cite{Farutin:21}, based on a point-particle model where advection is discarded in the vicinity of the particle, we find that bulk absorption `regularises' the pitchfork bifurcation, \textit{viz}., the speed initially grows away from the bifurcation like a square-root function rather than linearly as in the canonical isotropic scenario. Since our analysis is limited to weak bulk absorption, the regularisation we find does not fundamentally modify the structure of the weakly nonlinear analysis. In particular, the remote region remains essential and as a consequence the bifurcation still appears `singular' for $\sqrt{\mathrm{Da}_b}\ll\Pen-4\ll1$. Stronger bulk absorption would eliminate the remote region.  

\section{Concluding remarks}
\label{sec:conclusions}
Our main contribution in this part has been to identify an adjoint differential operator and auxiliary conditions that facilitate the derivation of nonlinear amplitude equations governing the steady velocity of a chemically active particle near the threshold for spontaneous motion. Our adjoint method circumvents the need to directly solve the inhomogeneous problem at quadratic order of a weakly nonlinear expansion, valid as the threshold is approached, making it relatively straightforward to analyze a wide range of perturbation scenarios; the technical simplification stems from the fact that the only problems that need to be explicitly solved, namely the linearised homogeneous problem at the threshold and its adjoint, are axisymmetric (about an arbitrary direction) and common to all scenarios, whereas the inhomogeneous problem at quadratic order is generally not axisymmetric and scenario-specific. To illustrate our approach, we have derived and then solved steady amplitude equations for a number of perturbation scenarios, demonstrating that sufficiently near the threshold weak perturbations can appreciably modify and enrich the landscape of steady solutions.

While we have introduced the adjoint homogeneous problem of Sec.~\ref{ssec:adjointproblem} as an auxiliary mathematical tool, we find it intriguing from a physical perspective that the transposed flow-solute coupling associated with that problem (cf.~\eqref{L star def}--\eqref{F T balances A}) can support spontaneous rectilinear motion of a particle, as represented by the adjoint homogeneous solutions \eqref{sol associated}. This may suggest an alternative physical mechanism for spontaneous self-propulsion, where a scalar field is associated with a body force on the fluid (hinting to buoyancy or electrostatics) and stress variations give rise to a surface flux of that scalar. As far as we know, such a mechanism for spontaneous motion has not yet been encountered.

We have only considered steady-state solutions in this part. As we shall see in the subsequent part, our adjoint method is equally useful, without modification, when studying the unsteady weakly nonlinear dynamics of a chemically active particle. This is because the weakly nonlinear dynamics evolve on a sufficiently long time scale such that the linear operator at first order of the weakly nonlinear expansion remains quasi-static (in fact, identical to that herein). Inspecting the steady problem formulation and form of the weakly nonlinear expansions in Sec.~\ref{sec:isotropic}, the appropriate long time scale can be deduced as $t_*=\epsilon^{-2}a_*^2/D_*$. With $t$ denoting time normalised by $t_*$, generalising the problem formulation of Sec.~\ref{ssec:formu} to allow for unsteadiness amounts to the addition of the term $\epsilon^2\partial{c}/\partial{t}$ to the advection-diffusion equation \eqref{c eq}; the particle-scale expansion remains quasi-steady at all relevant orders, while the leading-order remote-region equation \eqref{remote eq} becomes unsteady. As a consequence, the dynamics involve a history effect associated with the particle interacting with its own concentration wake. 
In the subsequent part, we shall develop unsteady nonlinear amplitude equations which include that  history effect. This will allow us to study the stability of the steady solutions found herein, explore transient dynamics, e.g., the alignment of the particle velocity vector with the direction of an external force field, as well as study inherently unsteady scenarios, such as: perturbation scenarios where there are no stable steady states  (as we shall see, this can occur in the case of non-uniform slip-coefficient perturbations, leading to stable circular motion); inter-particle and particle--wall interactions; and particles subjected to shear flow or an unsteady force field.

By suitably adapting the adjoint operators found here, it may be possible to develop weakly nonlinear theories for closely related scenarios where physico-chemical activity gives rise to spontaneous dynamics. In particular, it would be of interest to apply a similar approach to experimentally relevant models of so-called `solubilising' drops, whose activity can be modeled similarly to the canonical isotropic model, but with the flow typically being driven by a Marangoni, rather than diffusio-osmotic, effect \cite{Michelin:22}. Despite the different physics, we expect the form of the amplitude equations to be similar --- both the steady ones developed herein and the unsteady, history dependent, ones to be derived in the subsequent part --- such that solutions found for active particles could be adapted to such active drops.  

\textbf{Acknowledgments}. The author is grateful to Gunnar G.~Peng and Ehud Yariv for detailed  comments and Arianna Giunti for fruitful discussions. He also acknowledges the generous support of the Leverhulme Trust through Research Project Grant RPG-2021-161.

\appendix

\section{Surface differential operators}
\label{app:surfaceop}
Consider a surface that is locally covered by orthogonal curvilinear coordinates $(\nu_1,\nu_2)$, with associated unit vectors $(\hat{\bu}_1,\hat{\bu}_2)$ and scale factors $(h_1,h_2)$, such that $\partial{\br}/\partial\nu_1=h_1\hat{\bu}_1$ and $\partial{\br}/\partial\nu_2=h_2\hat{\bu}_2$, with the position vector $\br$ restricted to the surface and considered as a function of $\nu_1$ and $\nu_2$. We also define the normal unit vector $\bn=\hat{\bu}_1\times\hat{\bu}_2$. In some neighbourhood of the surface, $(\nu_1,\nu_2)$ can be extended to curvilinear bulk coordinates $(\nu_1,\nu_2,n)$, where $n$ is the distance from the surface along the normal such that $\br(\nu_1,\nu_2,n)=\br_s(\nu_1,\nu_2)+n\bn$, $\br_s(\nu_1,\nu_2)$ being the position on the surface with coordinates $(\nu_1,\nu_2)$. The extended  coordinates are clearly orthogonal for $n=0$; if $(\nu_1,\nu_2)$ trace lines of curvature of the surface, then the extended coordinates remain orthogonal  for $n\ne0$. 

The surface gradient of a scalar field, say $f$, can be defined as
\begin{equation}\label{surface gradient def}
\bnabla_sf=\frac{1}{h_1}\hat{\bu}_1\pd{f}{\nu_1}+\frac{1}{h_2}\hat{\bu}_2\pd{f}{\nu_2},
\end{equation}
with $f$ evaluated on the surface and considered as a function of $\nu_1$ and $\nu_2$. Comparing with the expression for the gradient operator in the extended curvilinear coordinates, we obtain the  coordinate-invariant relation $\bnabla_sf=(\tI-\bn\bn)\bcdot\bnabla f$ between the surface-gradient and gradient operators. In Sec.~\ref{ssec:properties}, we utilise a Stokes-type integral theorem saying that, for a closed surface, we have \cite{van:Book}
\begin{equation}
\oint\mathrm{d}A\,\bn\times \bnabla_sf=\bzero.
\end{equation}

The surface divergence of a vector field, say $\bP$, is defined as
\begin{equation}\label{sdiv def}
\bnabla_s\bcdot\bP=\frac{1}{h_1}\hat{\bu}_1\bcdot\pd{\bP}{\nu_1}+\frac{1}{h_2}\hat{\bu}_2\bcdot\pd{\bP}{\nu_2},
\end{equation}
with $\bP$ evaluated on the surface and considered as a function of $\nu_1$ and $\nu_2$; comparing with the expression for the divergence operator in the extended curvilinear coordinates, we see that $\bnabla_s\bcdot\bP=\left[(\tI-\bn\bn)\bcdot\bnabla\right]\bcdot\bP$. 
Let $\bP=\bP_{\parallel}+P_{\perp}\bn$, with $\bP_{\parallel}=(\tI-\bn\bn)\bcdot\bP$ and $P_{\perp}=\bP\bcdot\bn$. Then \eqref{sdiv def} gives 
\begin{equation}\label{surface def identitya}
\bnabla_s\bcdot\bP=\bnabla_s\bcdot\bP_{\parallel} +(\bnabla_s\bcdot\bn)P_{\perp}.
\end{equation}
For a tangential vector field $\bP_{\parallel}=P_1\hat{\bu}_1+P_2\hat{\bu}_2$, we find from \eqref{sdiv def}, using orthogonality and the definitions of the scale factors, the form 
\begin{equation}\label{sdiv form}
\bnabla_s\bcdot\bP_{\parallel}=\frac{1}{h_1h_2}\left\{\pd{}{\nu_1}\left(h_2P_1\right)+\pd{}{\nu_2}\left(h_1P_2\right)\right\}.
\end{equation} 

In Sec.~\ref{ssec:adjointoperator}, we employ two identities involving the surface-divergence operator. The first, 
\begin{equation}
\bnabla_s\bcdot(f\bP)=\bnabla_sf\bcdot\bP+f\bnabla_s\bcdot\bP,
\end{equation}
readily follows from \eqref{sdiv def}. In particular, in \eqref{manipulation} we have used this identity in the form $\bnabla_s\bcdot(f\bP_{\parallel})=\bnabla_sf\bcdot\bP+f\bnabla_s\bcdot\bP_{\parallel}$. The second states that, for a closed surface, we have  \cite{van:Book}
\begin{equation}
\oint\mathrm{d}A\,\bnabla_s\bcdot\bP=\oint\mathrm{d}A\,(\bnabla_s\bcdot\bn)P_{\perp}.
\end{equation}
In particular, tangential vector fields satisfy the Divergence-like law 
\begin{equation}
\oint\mathrm{d}A\,\bnabla_s\bcdot\bP_{\parallel}=0,
\end{equation}
which we have used in Sec.~\ref{ssec:adjointoperator} to carry out  integration by parts over the unit sphere.

Lastly, consider the case of spherical coordinates $(r,\theta,\phi)$, where $\theta$ is a polar angle and $\phi$ an azimuthal angle, with associated unit vectors $(\be_{\theta},\be_{\phi},\be_r)$.  Choosing our surface to be that of a sphere of radius $r_0$, we have $(\nu_1,\nu_2)=(\theta,\phi)$, $(\hat{\bu}_1,\hat{\bu}_2)=(\be_{\theta},\be_{\phi})$, $(h_{\theta},h_{\phi})=(r_0,r_0\sin\theta)$ and $\bn=\be_r$. From \eqref{surface gradient def}, we find 
\begin{equation}
\bnabla_sf=\frac{1}{r_0}\pd{f}{\theta}\be_{\theta}+\frac{1}{r_0\sin\theta}\pd{f}{\phi}\be_{\phi}, 
\end{equation}
which is needed in order to corroborate the solutions \eqref{homogeneous sol} to the homogeneous problem at linear order of the weakly nonlinear expansion, or, equivalently, the direct problem formulated in Sec.~\ref{ssec:homogeneousR}.
Writing $\bP=\bP_{\parallel}+P_{\perp}\be_r$, with $\bP_{\parallel}=P_{\theta}\be_{\theta}+P_{\phi}\be_{\phi}$, we find from \eqref{sdiv form} the expression
\begin{equation}\label{sc div tang}
\bnabla_s\bcdot\bP_{\parallel}=\frac{1}{r_0\sin\theta}\pd{}{\theta}\left(P_{\theta}\sin\theta\right)+\frac{1}{r_0\sin\theta}\pd{P_{\phi}}{\phi},
\end{equation}
which we use to solve the adjoint problem in Appendix \ref{app:spontaneousA}. The surface divergence of $\bP$ then follows from \eqref{surface def identitya}, with \eqref{sdiv def} giving $\bnabla_s\bcdot\bn=2/r_0$. 

\section{Adjoint spontaneous motion}\label{app:spontaneousA}
Consider the adjoint problem formulated in Sec.~\ref{ssec:adjointproblem} for the concentration field $c'$ and flow field $\bu'$, with associated pressure $p'$, stress tensor $\boldsymbol{\sigma}'$ and particle velocities $\bU'$ and $\boldsymbol{\Omega}'$. The problem consists of Laplace's equation \eqref{c eq A}, the concentration-coupled Stokes equations \eqref{u eqs A} and the natural auxiliary conditions associated with $\mathcal{L}^*$, namely the boundary conditions \eqref{bcs A}, the far-field conditions \eqref{far A} and the integral constraints \eqref{F T balances A}. Recall that the force is defined as in \eqref{F def new}; given that stress is not divergence-free in the adjoint problem, that definition differs from the conventional one (\ref{force torque}a). We shall constructively show that, like the direct problem defined in Sec.~\ref{ssec:homogeneousR} (and the order-$\epsilon$ homogeneous problem of Sec.~\ref{ssec:homogeneous}), this adjoint  problem possesses a family of non-trivial solutions describing steady rectilinear motion of the particle with an arbitrary particle velocity, without rotation. 

Let $\theta$ be the angle between $\br$ and $\bU'$. We introduce spherical coordinates $(r,\theta,\phi)$, with $r$ defined as in the main text and $\phi$ an azimuthal angle, and associated unit vectors $(\be_r,\be_{\theta},\be_{\phi})$. We look for solutions with the same angular dependence as the solutions of the direct problem (cf.~\eqref{linear sol}). Thus, we write
\refstepcounter{equation}
$$
\label{adjoint sol form}
c' = |\mathbf{U}'|\hat{c}(r)\cos\theta, \quad p'=|\mathbf{U}'|\hat{p}(r)\cos\theta, \quad \bu'=|\mathbf{U}'|\left\{\be_r\hat{u}(r)\cos\theta+\be_{\theta}\hat{v}(r)\sin\theta\right\},
\eqno{(\theequation a\!\!-\!\!c)}
$$
with $\mathbf{U}'$ an arbitrary vector and $\boldsymbol{\Omega}'=\bzero$. In terms of the reduced fields $\hat{c}(r)$, $\hat{u}(r)$ and $\hat{v}(r)$, the boundary conditions \eqref{bcs A} read as
\refstepcounter{equation}
$$
\label{adjoint sol bcs}
\frac{d\hat{c}}{dr}=2\frac{d\hat{v}}{dr}, \quad \hat{u}=0, \quad \hat{v}=0 \quad \text{at} \quad r=1
\eqno{(\theequation a\!\!-\!\!c)}
$$
and the far-field conditions \eqref{far A} read as
\refstepcounter{equation}
$$
\label{adjoint sol far}
\hat{c}=\mathcal{O}\left(\frac{1}{r^2}\right), \quad \hat{u}\to-1, \quad \hat{v}\to 1 \quad \text{as} \quad r\to\infty. 
\eqno{(\theequation a\!\!-\!\!c)}
$$
In (\ref{adjoint sol far}a), we used \eqref{sc div tang} to interpret the surface divergence of the tangential traction appearing in the adjoint boundary condition (\ref{bcs A}a). We also note that the decay condition (\ref{adjoint sol far}a), together with the fact that $c'$ satisfies Laplace's equation,  ensures that $\bnabla c' =\mathcal{O}(1/r^3)$, as required, in addition to (\ref{bcs A}a),  by the adjoint far-field condition (\ref{far A}a).

From Laplace's equation \eqref{c eq A}, the reduced field $\hat{c}(r)$ satisfies the differential equation
\begin{equation}\label{adjoint sol chat eq}
\frac{1}{r^2}\frac{d}{dr}\left(r^2\frac{d\hat{c}}{dr}\right)-\frac{2\hat{c}}{r^2}=0.
\end{equation}
Together with the decay condition (\ref{adjoint sol far}a), we find 
\begin{equation}\label{adjoint sol chat sol}
\hat{c}=\frac{\lambda}{r^2},
\end{equation}
where $\lambda$ is a constant to be determined. 

Given \eqref{adjoint sol chat sol}, the force density in the concentration-coupled momentum equation (\ref{u eqs A}b) is $\mathcal{O}(1/r^4)$ as $r\to\infty$. This suggests that a particular solution for the flow exists corresponding to a stress that is $\mathcal{O}(1/r^3)$ as $r\to\infty$. The contribution of such a particular solution to the force integral \eqref{F def new} thus vanishes. It follows that the zero-force constraint (\ref{F T balances A}a) can be satisfied by simply eliminating the Stokeslet term in the homogeneous solution to the concentration-coupled Stokes equations \eqref{u eqs A}. (Note however that the Stokeslet is no longer proportional to the conventional force on the particle as would be obtained by integrating the traction over the surface of the particle.) It follows that the force constraint can be represented by the following far-field condition on the reduced pressure $\hat{p}$ (taken to decay at infinity):
\begin{equation}\label{adjoint sol phat far}
\hat{p}=o\left(\frac{1}{r^2}\right) \quad \text{as} \quad r\to\infty. 
\end{equation}
Given the form of the solution \eqref{adjoint sol form}, the torque condition (\ref{F T balances A}b) is satisfied trivially.

Taking the divergence of (\ref{u eqs A}b), using (\ref{u eqs A}a), and substituting \eqref{adjoint sol chat sol}, we find that the pressure $p'$ satisfies a Poisson equation; in terms of the reduced pressure $\hat{p}$, we have
\begin{equation}
\frac{1}{r^2}\frac{d}{dr}\left(r^2\frac{d\hat{p}}{dr}\right)-\frac{2}{r^2}\hat{p}=-\frac{8\lambda}{r^5}.
\end{equation}
Solving in conjunction with \eqref{adjoint sol phat far}, we find 
\begin{equation}\label{adjoint solution phat sol}
\hat{p}=-\frac{2\lambda}{r^3}.
\end{equation}

Consider now the concentration-coupled momentum equation (\ref{u eqs A}b) in the radial and polar directions,
\refstepcounter{equation}
$$
\label{adjoint sol mom}
\frac{1}{r^2}\frac{d}{dr}\left(r^2\frac{d\hat{u}}{dr}\right)-\frac{4\hat{u}}{r^2}-\frac{4\hat{v}}{r^2}=\frac{d\hat{p}}{dr}-\frac{4\lambda}{r^4}, \quad \frac{1}{r^2}\frac{d}{dr}\left(r^2\frac{d\hat{v}}{dr}\right)-\frac{2\hat{v}}{r^2}-\frac{2\hat{u}}{r^2}=-\frac{\hat{p}}{r},
\eqno{(\theequation a,\!b)}
$$
along with the incompressibility constraint (\ref{u eqs A}a), 
\begin{equation}\label{adjoint sol sel}
\frac{1}{r^2}\frac{d}{dr}\left(r^2\hat{u}\right)+\frac{2}{r}\hat{v}=0.
\end{equation}
Substituting \eqref{adjoint sol sel} 
into (\ref{adjoint sol mom}a), we find
\begin{equation}\label{adjoint sol uhat eq}
\frac{d^2\hat{u}}{dr^2}+\frac{4}{r}\frac{d\hat{u}}{dr}=\frac{2\lambda}{r^4}.
\end{equation}
Solving \eqref{adjoint sol uhat eq} together with the boundary condition (\ref{adjoint sol bcs}b) and far-field condition (\ref{adjoint sol far}b) yields
\begin{equation}\label{adjoint sol uhat sol}
\hat{u}=-1-\frac{\lambda}{r^2}+\frac{1+\lambda}{r^3}. 
\end{equation}
We then readily find from the incompressibility condition \eqref{adjoint sol sel} that 
\begin{equation}
\hat{v}=1+\frac{1+\lambda}{2r^3},
\end{equation}
which together with \eqref{adjoint solution phat sol} and \eqref{adjoint sol uhat sol} trivially satisfies the polar momentum balance (\ref{adjoint sol mom}b). Finally, both of the  boundary conditions (\ref{adjoint sol bcs}a) and (\ref{adjoint sol bcs}c) are satisfied if 
\begin{equation}
\lambda=-3. 
\end{equation}

Using the geometric relations $\bU'\bcdot\be_r\be_r=|\bU'|\cos\theta\be_r$ and $\bU'\bcdot(\tI-\be_r\be_r)=-|\bU'|\sin\theta\be_{\theta}$, the above solution can be re-written in the cordinate-invariant form \eqref{sol associated} given in the main text.

\bibliography{refs}

\begin{thebibliography}{47}%
\makeatletter
\providecommand \@ifxundefined [1]{%
 \@ifx{#1\undefined}
}%
\providecommand \@ifnum [1]{%
 \ifnum #1\expandafter \@firstoftwo
 \else \expandafter \@secondoftwo
 \fi
}%
\providecommand \@ifx [1]{%
 \ifx #1\expandafter \@firstoftwo
 \else \expandafter \@secondoftwo
 \fi
}%
\providecommand \natexlab [1]{#1}%
\providecommand \enquote  [1]{``#1''}%
\providecommand \bibnamefont  [1]{#1}%
\providecommand \bibfnamefont [1]{#1}%
\providecommand \citenamefont [1]{#1}%
\providecommand \href@noop [0]{\@secondoftwo}%
\providecommand \href [0]{\begingroup \@sanitize@url \@href}%
\providecommand \@href[1]{\@@startlink{#1}\@@href}%
\providecommand \@@href[1]{\endgroup#1\@@endlink}%
\providecommand \@sanitize@url [0]{\catcode `\\12\catcode `\$12\catcode
  `\&12\catcode `\#12\catcode `\^12\catcode `\_12\catcode `\%12\relax}%
\providecommand \@@startlink[1]{}%
\providecommand \@@endlink[0]{}%
\providecommand \url  [0]{\begingroup\@sanitize@url \@url }%
\providecommand \@url [1]{\endgroup\@href {#1}{\urlprefix }}%
\providecommand \urlprefix  [0]{URL }%
\providecommand \Eprint [0]{\href }%
\providecommand \doibase [0]{http://dx.doi.org/}%
\providecommand \selectlanguage [0]{\@gobble}%
\providecommand \bibinfo  [0]{\@secondoftwo}%
\providecommand \bibfield  [0]{\@secondoftwo}%
\providecommand \translation [1]{[#1]}%
\providecommand \BibitemOpen [0]{}%
\providecommand \bibitemStop [0]{}%
\providecommand \bibitemNoStop [0]{.\EOS\space}%
\providecommand \EOS [0]{\spacefactor3000\relax}%
\providecommand \BibitemShut  [1]{\csname bibitem#1\endcsname}%
\let\auto@bib@innerbib\@empty
\bibitem [{\citenamefont {Anderson}(1989)}]{Anderson:89}%
  \BibitemOpen
  \bibfield  {author} {\bibinfo {author} {\bibfnamefont {J.~L.}\ \bibnamefont
  {Anderson}},\ }\bibfield  {title} {\enquote {\bibinfo {title} {Colloid
  transport by interfacial forces},}\ }\href@noop {} {\bibfield  {journal}
  {\bibinfo  {journal} {Annu. Rev. Fluid Mech.}\ }\textbf {\bibinfo {volume}
  {30}},\ \bibinfo {pages} {139--165} (\bibinfo {year} {1989})}\BibitemShut
  {NoStop}%
\bibitem [{\citenamefont {Golestanian}\ \emph {et~al.}(2005)\citenamefont
  {Golestanian}, \citenamefont {Liverpool},\ and\ \citenamefont
  {Ajdari}}]{Golestanian:05}%
  \BibitemOpen
  \bibfield  {author} {\bibinfo {author} {\bibfnamefont {R.}~\bibnamefont
  {Golestanian}}, \bibinfo {author} {\bibfnamefont {T.~B.}\ \bibnamefont
  {Liverpool}}, \ and\ \bibinfo {author} {\bibfnamefont {A.}~\bibnamefont
  {Ajdari}},\ }\bibfield  {title} {\enquote {\bibinfo {title} {Propulsion of a
  molecular machine by asymmetric distribution of reaction products},}\
  }\href@noop {} {\bibfield  {journal} {\bibinfo  {journal} {Phys. Rev. Lett.}\
  }\textbf {\bibinfo {volume} {94}},\ \bibinfo {pages} {220801} (\bibinfo
  {year} {2005})}\BibitemShut {NoStop}%
\bibitem [{\citenamefont {Golestanian}\ \emph {et~al.}(2007)\citenamefont
  {Golestanian}, \citenamefont {Liverpool},\ and\ \citenamefont
  {Ajdari}}]{Golestanian:07}%
  \BibitemOpen
  \bibfield  {author} {\bibinfo {author} {\bibfnamefont {R.}~\bibnamefont
  {Golestanian}}, \bibinfo {author} {\bibfnamefont {T.~B.}\ \bibnamefont
  {Liverpool}}, \ and\ \bibinfo {author} {\bibfnamefont {A.}~\bibnamefont
  {Ajdari}},\ }\bibfield  {title} {\enquote {\bibinfo {title} {Designing
  phoretic micro-and nano-swimmers},}\ }\href@noop {} {\bibfield  {journal}
  {\bibinfo  {journal} {New J. Phys.}\ }\textbf {\bibinfo {volume} {9}},\
  \bibinfo {pages} {126} (\bibinfo {year} {2007})}\BibitemShut {NoStop}%
\bibitem [{\citenamefont {Ebbens}\ \emph {et~al.}(2014)\citenamefont {Ebbens},
  \citenamefont {Gregory}, \citenamefont {Dunderdale}, \citenamefont {Howse},
  \citenamefont {Ibrahim}, \citenamefont {Liverpool},\ and\ \citenamefont
  {Golestanian}}]{Ebbens:14}%
  \BibitemOpen
  \bibfield  {author} {\bibinfo {author} {\bibfnamefont {S.}~\bibnamefont
  {Ebbens}}, \bibinfo {author} {\bibfnamefont {D.~A.}\ \bibnamefont {Gregory}},
  \bibinfo {author} {\bibfnamefont {G.}~\bibnamefont {Dunderdale}}, \bibinfo
  {author} {\bibfnamefont {J.~R.}\ \bibnamefont {Howse}}, \bibinfo {author}
  {\bibfnamefont {Y.}~\bibnamefont {Ibrahim}}, \bibinfo {author} {\bibfnamefont
  {T.~B.}\ \bibnamefont {Liverpool}}, \ and\ \bibinfo {author} {\bibfnamefont
  {R.}~\bibnamefont {Golestanian}},\ }\bibfield  {title} {\enquote {\bibinfo
  {title} {Electrokinetic effects in catalytic platinum-insulator janus
  swimmers},}\ }\href@noop {} {\bibfield  {journal} {\bibinfo  {journal} {EPL}\
  }\textbf {\bibinfo {volume} {106}},\ \bibinfo {pages} {058003} (\bibinfo
  {year} {2014})}\BibitemShut {NoStop}%
\bibitem [{\citenamefont {Michelin}\ and\ \citenamefont
  {Lauga}(2015)}]{Michelin:15}%
  \BibitemOpen
  \bibfield  {author} {\bibinfo {author} {\bibfnamefont {S.}~\bibnamefont
  {Michelin}}\ and\ \bibinfo {author} {\bibfnamefont {E.}~\bibnamefont
  {Lauga}},\ }\bibfield  {title} {\enquote {\bibinfo {title} {Autophoretic
  locomotion from geometric asymmetry},}\ }\href@noop {} {\bibfield  {journal}
  {\bibinfo  {journal} {Eur. Phys. J. E Soft Matter}\ }\textbf {\bibinfo
  {volume} {38}},\ \bibinfo {pages} {1--16} (\bibinfo {year}
  {2015})}\BibitemShut {NoStop}%
\bibitem [{\citenamefont {Popescu}\ \emph {et~al.}(2016)\citenamefont
  {Popescu}, \citenamefont {Uspal},\ and\ \citenamefont
  {Dietrich}}]{Popescu:16}%
  \BibitemOpen
  \bibfield  {author} {\bibinfo {author} {\bibfnamefont {M.~N.}\ \bibnamefont
  {Popescu}}, \bibinfo {author} {\bibfnamefont {W.~E.}\ \bibnamefont {Uspal}},
  \ and\ \bibinfo {author} {\bibfnamefont {S.}~\bibnamefont {Dietrich}},\
  }\bibfield  {title} {\enquote {\bibinfo {title} {Self-diffusiophoresis of
  chemically active colloids},}\ }\href@noop {} {\bibfield  {journal} {\bibinfo
   {journal} {Eur. Phys. J. Spec. Top.}\ }\textbf {\bibinfo {volume} {225}},\
  \bibinfo {pages} {2189--2206} (\bibinfo {year} {2016})}\BibitemShut {NoStop}%
\bibitem [{\citenamefont {Moran}\ and\ \citenamefont
  {Posner}(2017)}]{Moran:17}%
  \BibitemOpen
  \bibfield  {author} {\bibinfo {author} {\bibfnamefont {J.~L.}\ \bibnamefont
  {Moran}}\ and\ \bibinfo {author} {\bibfnamefont {J.~D.}\ \bibnamefont
  {Posner}},\ }\bibfield  {title} {\enquote {\bibinfo {title} {Phoretic
  self-propulsion},}\ }\href@noop {} {\bibfield  {journal} {\bibinfo  {journal}
  {Annu. Rev. Fluid Mech.}\ }\textbf {\bibinfo {volume} {49}},\ \bibinfo
  {pages} {511--540} (\bibinfo {year} {2017})}\BibitemShut {NoStop}%
\bibitem [{\citenamefont {Michelin}\ and\ \citenamefont
  {Lauga}(2017)}]{Michelin:17}%
  \BibitemOpen
  \bibfield  {author} {\bibinfo {author} {\bibfnamefont {S.}~\bibnamefont
  {Michelin}}\ and\ \bibinfo {author} {\bibfnamefont {E.}~\bibnamefont
  {Lauga}},\ }\bibfield  {title} {\enquote {\bibinfo {title} {Geometric tuning
  of self-propulsion for {Janus} catalytic particles},}\ }\href@noop {}
  {\bibfield  {journal} {\bibinfo  {journal} {Sci. Rep.}\ }\textbf {\bibinfo
  {volume} {7}},\ \bibinfo {pages} {42264} (\bibinfo {year}
  {2017})}\BibitemShut {NoStop}%
\bibitem [{\citenamefont {Rubinstein}\ \emph {et~al.}(2008)\citenamefont
  {Rubinstein}, \citenamefont {Manukyan}, \citenamefont {Staicu}, \citenamefont
  {Rubinstein}, \citenamefont {Zaltzman}, \citenamefont {Lammertink},
  \citenamefont {Mugele},\ and\ \citenamefont {Wessling}}]{Rubinstein:08}%
  \BibitemOpen
  \bibfield  {author} {\bibinfo {author} {\bibfnamefont {S.~M.}\ \bibnamefont
  {Rubinstein}}, \bibinfo {author} {\bibfnamefont {G.}~\bibnamefont
  {Manukyan}}, \bibinfo {author} {\bibfnamefont {A.}~\bibnamefont {Staicu}},
  \bibinfo {author} {\bibfnamefont {I.}~\bibnamefont {Rubinstein}}, \bibinfo
  {author} {\bibfnamefont {B.}~\bibnamefont {Zaltzman}}, \bibinfo {author}
  {\bibfnamefont {R.~G.~H.}\ \bibnamefont {Lammertink}}, \bibinfo {author}
  {\bibfnamefont {F.}~\bibnamefont {Mugele}}, \ and\ \bibinfo {author}
  {\bibfnamefont {M.}~\bibnamefont {Wessling}},\ }\bibfield  {title} {\enquote
  {\bibinfo {title} {Direct observation of a nonequilibrium electro-osmotic
  instability},}\ }\href@noop {} {\bibfield  {journal} {\bibinfo  {journal}
  {Phys. Rev. Lett.}\ }\textbf {\bibinfo {volume} {101}},\ \bibinfo {pages}
  {236101} (\bibinfo {year} {2008})}\BibitemShut {NoStop}%
\bibitem [{\citenamefont {Game}\ \emph {et~al.}(2017)\citenamefont {Game},
  \citenamefont {Hodes}, \citenamefont {Keaveny},\ and\ \citenamefont
  {Papageorgiou}}]{Game:17}%
  \BibitemOpen
  \bibfield  {author} {\bibinfo {author} {\bibfnamefont {S.~E.}\ \bibnamefont
  {Game}}, \bibinfo {author} {\bibfnamefont {M.}~\bibnamefont {Hodes}},
  \bibinfo {author} {\bibfnamefont {E.~E.}\ \bibnamefont {Keaveny}}, \ and\
  \bibinfo {author} {\bibfnamefont {D.~T.}\ \bibnamefont {Papageorgiou}},\
  }\bibfield  {title} {\enquote {\bibinfo {title} {Physical mechanisms relevant
  to flow resistance in textured microchannels},}\ }\href@noop {} {\bibfield
  {journal} {\bibinfo  {journal} {Phys. Rev. Fluids}\ }\textbf {\bibinfo
  {volume} {2}},\ \bibinfo {pages} {094102--23} (\bibinfo {year}
  {2017})}\BibitemShut {NoStop}%
\bibitem [{\citenamefont {Chen}\ \emph {et~al.}(2021)\citenamefont {Chen},
  \citenamefont {Chong}, \citenamefont {Liu}, \citenamefont {Verzicco},\ and\
  \citenamefont {Lohse}}]{Chen:21}%
  \BibitemOpen
  \bibfield  {author} {\bibinfo {author} {\bibfnamefont {Y.}~\bibnamefont
  {Chen}}, \bibinfo {author} {\bibfnamefont {K.~L.}\ \bibnamefont {Chong}},
  \bibinfo {author} {\bibfnamefont {L.}~\bibnamefont {Liu}}, \bibinfo {author}
  {\bibfnamefont {R.}~\bibnamefont {Verzicco}}, \ and\ \bibinfo {author}
  {\bibfnamefont {D.}~\bibnamefont {Lohse}},\ }\bibfield  {title} {\enquote
  {\bibinfo {title} {Instabilities driven by diffusiophoretic flow on catalytic
  surfaces},}\ }\href@noop {} {\bibfield  {journal} {\bibinfo  {journal} {J.
  Fluid Mech.}\ }\textbf {\bibinfo {volume} {919}},\ \bibinfo {pages} {A10}
  (\bibinfo {year} {2021})}\BibitemShut {NoStop}%
\bibitem [{\citenamefont {Michelin}\ \emph {et~al.}(2013)\citenamefont
  {Michelin}, \citenamefont {Lauga},\ and\ \citenamefont
  {Bartolo}}]{Michelin:13}%
  \BibitemOpen
  \bibfield  {author} {\bibinfo {author} {\bibfnamefont {S.}~\bibnamefont
  {Michelin}}, \bibinfo {author} {\bibfnamefont {E.}~\bibnamefont {Lauga}}, \
  and\ \bibinfo {author} {\bibfnamefont {D.}~\bibnamefont {Bartolo}},\
  }\bibfield  {title} {\enquote {\bibinfo {title} {Spontaneous autophoretic
  motion of isotropic particles},}\ }\href@noop {} {\bibfield  {journal}
  {\bibinfo  {journal} {Phys. Fluids}\ }\textbf {\bibinfo {volume} {25}},\
  \bibinfo {pages} {061701} (\bibinfo {year} {2013})}\BibitemShut {NoStop}%
\bibitem [{\citenamefont {Michelin}\ and\ \citenamefont
  {Lauga}(2014)}]{Michelin:14}%
  \BibitemOpen
  \bibfield  {author} {\bibinfo {author} {\bibfnamefont {S.}~\bibnamefont
  {Michelin}}\ and\ \bibinfo {author} {\bibfnamefont {E.}~\bibnamefont
  {Lauga}},\ }\bibfield  {title} {\enquote {\bibinfo {title} {Phoretic
  self-propulsion at finite {P\'eclet} numbers},}\ }\href@noop {} {\bibfield
  {journal} {\bibinfo  {journal} {J. Fluid Mech.}\ }\textbf {\bibinfo {volume}
  {747}},\ \bibinfo {pages} {572--604} (\bibinfo {year} {2014})}\BibitemShut
  {NoStop}%
\bibitem [{\citenamefont {Morozov}\ and\ \citenamefont
  {Michelin}(2019{\natexlab{a}})}]{Morozov:19}%
  \BibitemOpen
  \bibfield  {author} {\bibinfo {author} {\bibfnamefont {M.}~\bibnamefont
  {Morozov}}\ and\ \bibinfo {author} {\bibfnamefont {S.}~\bibnamefont
  {Michelin}},\ }\bibfield  {title} {\enquote {\bibinfo {title} {Nonlinear
  dynamics of a chemically-active drop: From steady to chaotic
  self-propulsion},}\ }\href@noop {} {\bibfield  {journal} {\bibinfo  {journal}
  {J. Chem. Phys.}\ }\textbf {\bibinfo {volume} {150}},\ \bibinfo {pages}
  {044110} (\bibinfo {year} {2019}{\natexlab{a}})}\BibitemShut {NoStop}%
\bibitem [{\citenamefont {Kailasham}\ and\ \citenamefont
  {Khair}(2022)}]{Kailasham:22}%
  \BibitemOpen
  \bibfield  {author} {\bibinfo {author} {\bibfnamefont {R.}~\bibnamefont
  {Kailasham}}\ and\ \bibinfo {author} {\bibfnamefont {A.~S.}\ \bibnamefont
  {Khair}},\ }\bibfield  {title} {\enquote {\bibinfo {title} {Dynamics of
  forced and unforced autophoretic particles},}\ }\href@noop {} {\bibfield
  {journal} {\bibinfo  {journal} {J. Fluid Mech.}\ }\textbf {\bibinfo {volume}
  {948}},\ \bibinfo {pages} {A41} (\bibinfo {year} {2022})}\BibitemShut
  {NoStop}%
\bibitem [{\citenamefont {Rednikov}\ \emph {et~al.}(1994)\citenamefont
  {Rednikov}, \citenamefont {Ryazantsev},\ and\ \citenamefont
  {Velarde}}]{Rednikov:94}%
  \BibitemOpen
  \bibfield  {author} {\bibinfo {author} {\bibfnamefont {A.~Y.}\ \bibnamefont
  {Rednikov}}, \bibinfo {author} {\bibfnamefont {Y.~S.}\ \bibnamefont
  {Ryazantsev}}, \ and\ \bibinfo {author} {\bibfnamefont {M.~G.}\ \bibnamefont
  {Velarde}},\ }\bibfield  {title} {\enquote {\bibinfo {title} {Drop motion
  with surfactant transfer in a homogeneous surrounding},}\ }\href@noop {}
  {\bibfield  {journal} {\bibinfo  {journal} {Phys. Fluids}\ }\textbf {\bibinfo
  {volume} {6}},\ \bibinfo {pages} {451--468} (\bibinfo {year}
  {1994})}\BibitemShut {NoStop}%
\bibitem [{\citenamefont {Schmitt}\ and\ \citenamefont
  {Stark}(2013)}]{Schmitt:13}%
  \BibitemOpen
  \bibfield  {author} {\bibinfo {author} {\bibfnamefont {M.}~\bibnamefont
  {Schmitt}}\ and\ \bibinfo {author} {\bibfnamefont {H.}~\bibnamefont
  {Stark}},\ }\bibfield  {title} {\enquote {\bibinfo {title} {Swimming active
  droplet: A theoretical analysis},}\ }\href@noop {} {\bibfield  {journal}
  {\bibinfo  {journal} {EPL}\ }\textbf {\bibinfo {volume} {101}},\ \bibinfo
  {pages} {44008} (\bibinfo {year} {2013})}\BibitemShut {NoStop}%
\bibitem [{\citenamefont {Izri}\ \emph {et~al.}(2014)\citenamefont {Izri},
  \citenamefont {Van Der~Linden}, \citenamefont {Michelin},\ and\ \citenamefont
  {Dauchot}}]{Izri:14}%
  \BibitemOpen
  \bibfield  {author} {\bibinfo {author} {\bibfnamefont {Z.}~\bibnamefont
  {Izri}}, \bibinfo {author} {\bibfnamefont {M.~N.}\ \bibnamefont {Van
  Der~Linden}}, \bibinfo {author} {\bibfnamefont {S.}~\bibnamefont {Michelin}},
  \ and\ \bibinfo {author} {\bibfnamefont {O.}~\bibnamefont {Dauchot}},\
  }\bibfield  {title} {\enquote {\bibinfo {title} {Self-propulsion of pure
  water droplets by spontaneous {Marangoni}-stress-driven motion},}\
  }\href@noop {} {\bibfield  {journal} {\bibinfo  {journal} {Phys. Rev. Lett.}\
  }\textbf {\bibinfo {volume} {113}},\ \bibinfo {pages} {248302} (\bibinfo
  {year} {2014})}\BibitemShut {NoStop}%
\bibitem [{\citenamefont {Suda}\ \emph {et~al.}(2021)\citenamefont {Suda},
  \citenamefont {Suda}, \citenamefont {Ohmura},\ and\ \citenamefont
  {Ichikawa}}]{Suda:21}%
  \BibitemOpen
  \bibfield  {author} {\bibinfo {author} {\bibfnamefont {S.}~\bibnamefont
  {Suda}}, \bibinfo {author} {\bibfnamefont {T.}~\bibnamefont {Suda}}, \bibinfo
  {author} {\bibfnamefont {T.}~\bibnamefont {Ohmura}}, \ and\ \bibinfo {author}
  {\bibfnamefont {M.}~\bibnamefont {Ichikawa}},\ }\bibfield  {title} {\enquote
  {\bibinfo {title} {Straight-to-curvilinear motion transition of a swimming
  droplet caused by the susceptibility to fluctuations},}\ }\href@noop {}
  {\bibfield  {journal} {\bibinfo  {journal} {Phys. Rev. Lett.}\ }\textbf
  {\bibinfo {volume} {127}},\ \bibinfo {pages} {088005} (\bibinfo {year}
  {2021})}\BibitemShut {NoStop}%
\bibitem [{\citenamefont {Hokmabad}\ \emph {et~al.}(2021)\citenamefont
  {Hokmabad}, \citenamefont {Dey}, \citenamefont {Jalaal}, \citenamefont
  {Mohanty}, \citenamefont {Almukambetova}, \citenamefont {Baldwin},
  \citenamefont {Lohse},\ and\ \citenamefont {Maass}}]{Hokmabad:21}%
  \BibitemOpen
  \bibfield  {author} {\bibinfo {author} {\bibfnamefont {B.~V.}\ \bibnamefont
  {Hokmabad}}, \bibinfo {author} {\bibfnamefont {R.}~\bibnamefont {Dey}},
  \bibinfo {author} {\bibfnamefont {M.}~\bibnamefont {Jalaal}}, \bibinfo
  {author} {\bibfnamefont {D.}~\bibnamefont {Mohanty}}, \bibinfo {author}
  {\bibfnamefont {M.}~\bibnamefont {Almukambetova}}, \bibinfo {author}
  {\bibfnamefont {K.~A.}\ \bibnamefont {Baldwin}}, \bibinfo {author}
  {\bibfnamefont {D.}~\bibnamefont {Lohse}}, \ and\ \bibinfo {author}
  {\bibfnamefont {C.~C.}\ \bibnamefont {Maass}},\ }\bibfield  {title} {\enquote
  {\bibinfo {title} {Emergence of bimodal motility in active droplets},}\
  }\href@noop {} {\bibfield  {journal} {\bibinfo  {journal} {Phys. Rev. X}\
  }\textbf {\bibinfo {volume} {11}},\ \bibinfo {pages} {011043} (\bibinfo
  {year} {2021})}\BibitemShut {NoStop}%
\bibitem [{\citenamefont {Li}(2022)}]{Li:22}%
  \BibitemOpen
  \bibfield  {author} {\bibinfo {author} {\bibfnamefont {Gaojin}\ \bibnamefont
  {Li}},\ }\bibfield  {title} {\enquote {\bibinfo {title} {Swimming dynamics of
  a self-propelled droplet},}\ }\href {\doibase 10.1017/jfm.2021.1154}
  {\bibfield  {journal} {\bibinfo  {journal} {J. Fluid Mech.}\ }\textbf
  {\bibinfo {volume} {934}},\ \bibinfo {pages} {A20} (\bibinfo {year}
  {2022})}\BibitemShut {NoStop}%
\bibitem [{\citenamefont {Hokmabad}\ \emph {et~al.}(2022)\citenamefont
  {Hokmabad}, \citenamefont {Nishide}, \citenamefont {Ramesh},\ and\
  \citenamefont {Maass}}]{Hokmabad:21b}%
  \BibitemOpen
  \bibfield  {author} {\bibinfo {author} {\bibfnamefont {B.~V.}\ \bibnamefont
  {Hokmabad}}, \bibinfo {author} {\bibfnamefont {A.}~\bibnamefont {Nishide}},
  \bibinfo {author} {\bibfnamefont {P.}~\bibnamefont {Ramesh}}, \ and\ \bibinfo
  {author} {\bibfnamefont {C.~C.}\ \bibnamefont {Maass}},\ }\bibfield  {title}
  {\enquote {\bibinfo {title} {Spontaneously rotating clusters of active
  droplets},}\ }\href@noop {} {\bibfield  {journal} {\bibinfo  {journal} {Soft
  matter}\ }\textbf {\bibinfo {volume} {18}},\ \bibinfo {pages} {2731--2741}
  (\bibinfo {year} {2022})}\BibitemShut {NoStop}%
\bibitem [{\citenamefont {Michelin}(2022)}]{Michelin:22}%
  \BibitemOpen
  \bibfield  {author} {\bibinfo {author} {\bibfnamefont {S.}~\bibnamefont
  {Michelin}},\ }\bibfield  {title} {\enquote {\bibinfo {title}
  {Self-propulsion of chemically active droplets},}\ }\href@noop {} {\bibfield
  {journal} {\bibinfo  {journal} {Ann. Rev. Fluid Mech.}\ }\textbf {\bibinfo
  {volume} {55}} (\bibinfo {year} {2022})}\BibitemShut {NoStop}%
\bibitem [{\citenamefont {Farutin}\ and\ \citenamefont
  {Misbah}(2021)}]{Farutin:21}%
  \BibitemOpen
  \bibfield  {author} {\bibinfo {author} {\bibfnamefont {A.}~\bibnamefont
  {Farutin}}\ and\ \bibinfo {author} {\bibfnamefont {C.}~\bibnamefont
  {Misbah}},\ }\bibfield  {title} {\enquote {\bibinfo {title} {Singular
  bifurcations: a regularization theory},}\ }\href@noop {} {\bibfield
  {journal} {\bibinfo  {journal} {arXiv preprint arXiv:2112.12094}\ } (\bibinfo
  {year} {2021})}\BibitemShut {NoStop}%
\bibitem [{\citenamefont {Saha}\ and\ \citenamefont {Yariv}(2022)}]{Saha:22}%
  \BibitemOpen
  \bibfield  {author} {\bibinfo {author} {\bibfnamefont {S.}~\bibnamefont
  {Saha}}\ and\ \bibinfo {author} {\bibfnamefont {E.}~\bibnamefont {Yariv}},\
  }\bibfield  {title} {\enquote {\bibinfo {title} {Phoretic self-propulsion of
  a slightly inhomogeneous disc},}\ }\href@noop {} {\bibfield  {journal}
  {\bibinfo  {journal} {J. Fluid Mech.}\ }\textbf {\bibinfo {volume} {940}}
  (\bibinfo {year} {2022})}\BibitemShut {NoStop}%
\bibitem [{\citenamefont {Picella}\ and\ \citenamefont
  {Michelin}(2022)}]{Picella:22}%
  \BibitemOpen
  \bibfield  {author} {\bibinfo {author} {\bibfnamefont {F.}~\bibnamefont
  {Picella}}\ and\ \bibinfo {author} {\bibfnamefont {S.}~\bibnamefont
  {Michelin}},\ }\bibfield  {title} {\enquote {\bibinfo {title} {Confined
  self-propulsion of an isotropic active colloid},}\ }\href@noop {} {\bibfield
  {journal} {\bibinfo  {journal} {J. Fluid Mech.}\ }\textbf {\bibinfo {volume}
  {933}} (\bibinfo {year} {2022})}\BibitemShut {NoStop}%
\bibitem [{\citenamefont {Yariv}\ and\ \citenamefont
  {Kaynan}(2017)}]{Yariv:17}%
  \BibitemOpen
  \bibfield  {author} {\bibinfo {author} {\bibfnamefont {E.}~\bibnamefont
  {Yariv}}\ and\ \bibinfo {author} {\bibfnamefont {U.}~\bibnamefont {Kaynan}},\
  }\bibfield  {title} {\enquote {\bibinfo {title} {Phoretic drag reduction of
  chemically active homogeneous spheres under force fields and shear flows},}\
  }\href@noop {} {\bibfield  {journal} {\bibinfo  {journal} {Phys. Rev.
  Fluids}\ }\textbf {\bibinfo {volume} {2}},\ \bibinfo {pages} {012201}
  (\bibinfo {year} {2017})}\BibitemShut {NoStop}%
\bibitem [{\citenamefont {Saha}\ \emph {et~al.}(2021)\citenamefont {Saha},
  \citenamefont {Yariv},\ and\ \citenamefont {Schnitzer}}]{Saha:21}%
  \BibitemOpen
  \bibfield  {author} {\bibinfo {author} {\bibfnamefont {S.}~\bibnamefont
  {Saha}}, \bibinfo {author} {\bibfnamefont {E.}~\bibnamefont {Yariv}}, \ and\
  \bibinfo {author} {\bibfnamefont {O.}~\bibnamefont {Schnitzer}},\ }\bibfield
  {title} {\enquote {\bibinfo {title} {Isotropically active colloids under
  uniform force fields: from forced to spontaneous motion},}\ }\href@noop {}
  {\bibfield  {journal} {\bibinfo  {journal} {J. Fluid Mech.}\ }\textbf
  {\bibinfo {volume} {916}} (\bibinfo {year} {2021})}\BibitemShut {NoStop}%
\bibitem [{\citenamefont {Riazantsev}\ and\ \citenamefont
  {Rednikov}(1992)}]{Riazantsev:92}%
  \BibitemOpen
  \bibfield  {author} {\bibinfo {author} {\bibfnamefont {I.~S.}\ \bibnamefont
  {Riazantsev}}\ and\ \bibinfo {author} {\bibfnamefont {A.~E.}\ \bibnamefont
  {Rednikov}},\ }\bibfield  {title} {\enquote {\bibinfo {title} {Self-sustained
  motion of a drop in homogeneous surroundings},}\ }in\ \href@noop {} {\emph
  {\bibinfo {booktitle} {Washington, DC International Astronautical Federation
  Congress}}}\ (\bibinfo {year} {1992})\BibitemShut {NoStop}%
\bibitem [{\citenamefont {Morozov}\ and\ \citenamefont
  {Michelin}(2019{\natexlab{b}})}]{Morozov:19b}%
  \BibitemOpen
  \bibfield  {author} {\bibinfo {author} {\bibfnamefont {M.}~\bibnamefont
  {Morozov}}\ and\ \bibinfo {author} {\bibfnamefont {S.}~\bibnamefont
  {Michelin}},\ }\bibfield  {title} {\enquote {\bibinfo {title}
  {Self-propulsion near the onset of marangoni instability of deformable active
  droplets},}\ }\href@noop {} {\bibfield  {journal} {\bibinfo  {journal} {J.
  Fluid Mech.}\ }\textbf {\bibinfo {volume} {860}},\ \bibinfo {pages}
  {711--738} (\bibinfo {year} {2019}{\natexlab{b}})}\BibitemShut {NoStop}%
\bibitem [{\citenamefont {Rednikov}\ \emph {et~al.}(1995)\citenamefont
  {Rednikov}, \citenamefont {Kurdyumov}, \citenamefont {Ryazantsev},\ and\
  \citenamefont {Velarde}}]{Rednikov:95}%
  \BibitemOpen
  \bibfield  {author} {\bibinfo {author} {\bibfnamefont {A.~Y.}\ \bibnamefont
  {Rednikov}}, \bibinfo {author} {\bibfnamefont {V.~N.}\ \bibnamefont
  {Kurdyumov}}, \bibinfo {author} {\bibfnamefont {Y.~S.}\ \bibnamefont
  {Ryazantsev}}, \ and\ \bibinfo {author} {\bibfnamefont {M.~G.}\ \bibnamefont
  {Velarde}},\ }\bibfield  {title} {\enquote {\bibinfo {title} {The role of
  time-varying gravity on the motion of a drop induced by marangoni
  instability},}\ }\href@noop {} {\bibfield  {journal} {\bibinfo  {journal}
  {Phys. Fluids}\ }\textbf {\bibinfo {volume} {7}},\ \bibinfo {pages}
  {2670--2678} (\bibinfo {year} {1995})}\BibitemShut {NoStop}%
\bibitem [{\citenamefont {Lippera}\ \emph
  {et~al.}(2020{\natexlab{a}})\citenamefont {Lippera}, \citenamefont {Morozov},
  \citenamefont {Benzaquen},\ and\ \citenamefont {Michelin}}]{Lippera:20}%
  \BibitemOpen
  \bibfield  {author} {\bibinfo {author} {\bibfnamefont {K.}~\bibnamefont
  {Lippera}}, \bibinfo {author} {\bibfnamefont {M.}~\bibnamefont {Morozov}},
  \bibinfo {author} {\bibfnamefont {M.}~\bibnamefont {Benzaquen}}, \ and\
  \bibinfo {author} {\bibfnamefont {S.}~\bibnamefont {Michelin}},\ }\bibfield
  {title} {\enquote {\bibinfo {title} {Collisions and rebounds of chemically
  active droplets},}\ }\href@noop {} {\bibfield  {journal} {\bibinfo  {journal}
  {J. Fluid Mech.}\ }\textbf {\bibinfo {volume} {886}},\ \bibinfo {pages}
  {1843--35} (\bibinfo {year} {2020}{\natexlab{a}})}\BibitemShut {NoStop}%
\bibitem [{\citenamefont {Lippera}\ \emph
  {et~al.}(2020{\natexlab{b}})\citenamefont {Lippera}, \citenamefont
  {Benzaquen},\ and\ \citenamefont {Michelin}}]{Lippera:20a}%
  \BibitemOpen
  \bibfield  {author} {\bibinfo {author} {\bibfnamefont {K.}~\bibnamefont
  {Lippera}}, \bibinfo {author} {\bibfnamefont {M.}~\bibnamefont {Benzaquen}},
  \ and\ \bibinfo {author} {\bibfnamefont {S.}~\bibnamefont {Michelin}},\
  }\bibfield  {title} {\enquote {\bibinfo {title} {Bouncing, chasing, or
  pausing: Asymmetric collisions of active droplets},}\ }\href@noop {}
  {\bibfield  {journal} {\bibinfo  {journal} {Phys. Rev. Fluids}\ }\textbf
  {\bibinfo {volume} {5}},\ \bibinfo {pages} {032201} (\bibinfo {year}
  {2020}{\natexlab{b}})}\BibitemShut {NoStop}%
\bibitem [{\citenamefont {Desai}\ and\ \citenamefont
  {Michelin}(2021)}]{Desai:21}%
  \BibitemOpen
  \bibfield  {author} {\bibinfo {author} {\bibfnamefont {N.}~\bibnamefont
  {Desai}}\ and\ \bibinfo {author} {\bibfnamefont {S.}~\bibnamefont
  {Michelin}},\ }\bibfield  {title} {\enquote {\bibinfo {title} {Instability
  and self-propulsion of active droplets along a wall},}\ }\href@noop {}
  {\bibfield  {journal} {\bibinfo  {journal} {Phys. Rev. Fluids}\ }\textbf
  {\bibinfo {volume} {6}},\ \bibinfo {pages} {114103} (\bibinfo {year}
  {2021})}\BibitemShut {NoStop}%
\bibitem [{\citenamefont {Hu}\ \emph {et~al.}(2019)\citenamefont {Hu},
  \citenamefont {Lin}, \citenamefont {Rafai},\ and\ \citenamefont
  {Misbah}}]{Hu:19}%
  \BibitemOpen
  \bibfield  {author} {\bibinfo {author} {\bibfnamefont {W.~F.}\ \bibnamefont
  {Hu}}, \bibinfo {author} {\bibfnamefont {T.~S.}\ \bibnamefont {Lin}},
  \bibinfo {author} {\bibfnamefont {S.}~\bibnamefont {Rafai}}, \ and\ \bibinfo
  {author} {\bibfnamefont {C.}~\bibnamefont {Misbah}},\ }\bibfield  {title}
  {\enquote {\bibinfo {title} {Chaotic swimming of phoretic particles},}\
  }\href@noop {} {\bibfield  {journal} {\bibinfo  {journal} {Phys. Rev. Lett.}\
  }\textbf {\bibinfo {volume} {123}},\ \bibinfo {pages} {238004} (\bibinfo
  {year} {2019})}\BibitemShut {NoStop}%
\bibitem [{\citenamefont {Farutin}\ \emph {et~al.}(2021)\citenamefont
  {Farutin}, \citenamefont {Rizvi}, \citenamefont {Hu}, \citenamefont {Lin},
  \citenamefont {Rafai},\ and\ \citenamefont {Misbah}}]{Farutin:21b}%
  \BibitemOpen
  \bibfield  {author} {\bibinfo {author} {\bibfnamefont {A.}~\bibnamefont
  {Farutin}}, \bibinfo {author} {\bibfnamefont {M.~S.}\ \bibnamefont {Rizvi}},
  \bibinfo {author} {\bibfnamefont {W.~F.}\ \bibnamefont {Hu}}, \bibinfo
  {author} {\bibfnamefont {T.~S.}\ \bibnamefont {Lin}}, \bibinfo {author}
  {\bibfnamefont {S.}~\bibnamefont {Rafai}}, \ and\ \bibinfo {author}
  {\bibfnamefont {C.}~\bibnamefont {Misbah}},\ }\bibfield  {title} {\enquote
  {\bibinfo {title} {A reduced model for a phoretic swimmer},}\ }\href@noop {}
  {\bibfield  {journal} {\bibinfo  {journal} {arXiv preprint arXiv:2112.12023}\
  } (\bibinfo {year} {2021})}\BibitemShut {NoStop}%
\bibitem [{\citenamefont {Hu}\ \emph {et~al.}(2022)\citenamefont {Hu},
  \citenamefont {Lin}, \citenamefont {Rafai},\ and\ \citenamefont
  {Misbah}}]{Hu:22}%
  \BibitemOpen
  \bibfield  {author} {\bibinfo {author} {\bibfnamefont {W.~F.}\ \bibnamefont
  {Hu}}, \bibinfo {author} {\bibfnamefont {T.~S.}\ \bibnamefont {Lin}},
  \bibinfo {author} {\bibfnamefont {S.}~\bibnamefont {Rafai}}, \ and\ \bibinfo
  {author} {\bibfnamefont {C.}~\bibnamefont {Misbah}},\ }\bibfield  {title}
  {\enquote {\bibinfo {title} {Spontaneous locomotion of phoretic particles in
  three dimensions},}\ }\href@noop {} {\bibfield  {journal} {\bibinfo
  {journal} {Phys. Rev. Fluids}\ }\textbf {\bibinfo {volume} {7}},\ \bibinfo
  {pages} {034003} (\bibinfo {year} {2022})}\BibitemShut {NoStop}%
\bibitem [{\citenamefont {Boniface}\ \emph {et~al.}(2019)\citenamefont
  {Boniface}, \citenamefont {Cottin-Bizonne}, \citenamefont {Kervil},
  \citenamefont {Ybert},\ and\ \citenamefont {Detcheverry}}]{Boniface:19}%
  \BibitemOpen
  \bibfield  {author} {\bibinfo {author} {\bibfnamefont {D.}~\bibnamefont
  {Boniface}}, \bibinfo {author} {\bibfnamefont {C.}~\bibnamefont
  {Cottin-Bizonne}}, \bibinfo {author} {\bibfnamefont {R.}~\bibnamefont
  {Kervil}}, \bibinfo {author} {\bibfnamefont {C.}~\bibnamefont {Ybert}}, \
  and\ \bibinfo {author} {\bibfnamefont {F.}~\bibnamefont {Detcheverry}},\
  }\bibfield  {title} {\enquote {\bibinfo {title} {Self-propulsion of symmetric
  chemically active particles: {Point-source} model and experiments on camphor
  disks},}\ }\href@noop {} {\bibfield  {journal} {\bibinfo  {journal} {Phys.
  Rev. E}\ }\textbf {\bibinfo {volume} {99}},\ \bibinfo {pages} {062605}
  (\bibinfo {year} {2019})}\BibitemShut {NoStop}%
\bibitem [{\citenamefont {Lippera}\ \emph
  {et~al.}(2020{\natexlab{c}})\citenamefont {Lippera}, \citenamefont
  {Benzaquen},\ and\ \citenamefont {Michelin}}]{Lippera:20b}%
  \BibitemOpen
  \bibfield  {author} {\bibinfo {author} {\bibfnamefont {K.}~\bibnamefont
  {Lippera}}, \bibinfo {author} {\bibfnamefont {M.}~\bibnamefont {Benzaquen}},
  \ and\ \bibinfo {author} {\bibfnamefont {S.}~\bibnamefont {Michelin}},\
  }\bibfield  {title} {\enquote {\bibinfo {title} {Alignment and scattering of
  colliding active droplets},}\ }\href@noop {} {\bibfield  {journal} {\bibinfo
  {journal} {Soft Matter}\ } (\bibinfo {year}
  {2020}{\natexlab{c}})}\BibitemShut {NoStop}%
\bibitem [{\citenamefont {Ye}\ and\ \citenamefont {Velarde}(1994)}]{Ye:94}%
  \BibitemOpen
  \bibfield  {author} {\bibinfo {author} {\bibfnamefont {Ryazantsev Y.~S.}\
  \bibnamefont {Ye}, \bibfnamefont {A.~R.}}\ and\ \bibinfo {author}
  {\bibfnamefont {M.~G.}\ \bibnamefont {Velarde}},\ }\bibfield  {title}
  {\enquote {\bibinfo {title} {Drop motion and the maragoni effect. interaction
  of modes},}\ }\href@noop {} {\bibfield  {journal} {\bibinfo  {journal} {Phys.
  Scr.}\ }\textbf {\bibinfo {volume} {1994}},\ \bibinfo {pages} {115} (\bibinfo
  {year} {1994})}\BibitemShut {NoStop}%
\bibitem [{\citenamefont {Acrivos}\ and\ \citenamefont
  {Taylor}(1962)}]{Acrivos:62}%
  \BibitemOpen
  \bibfield  {author} {\bibinfo {author} {\bibfnamefont {A.}~\bibnamefont
  {Acrivos}}\ and\ \bibinfo {author} {\bibfnamefont {T.~D.}\ \bibnamefont
  {Taylor}},\ }\bibfield  {title} {\enquote {\bibinfo {title} {Heat and mass
  transfer from single spheres in {Stokes} flow},}\ }\href@noop {} {\bibfield
  {journal} {\bibinfo  {journal} {Phys. Fluids}\ }\textbf {\bibinfo {volume}
  {5}},\ \bibinfo {pages} {387--394} (\bibinfo {year} {1962})}\BibitemShut
  {NoStop}%
\bibitem [{\citenamefont {Hinch}(1991)}]{Hinch:book}%
  \BibitemOpen
  \bibfield  {author} {\bibinfo {author} {\bibfnamefont {E.~J.}\ \bibnamefont
  {Hinch}},\ }\href@noop {} {\emph {\bibinfo {title} {Perturbation Methods}}}\
  (\bibinfo  {publisher} {Cambridge University Press},\ \bibinfo {address}
  {Cambridge},\ \bibinfo {year} {1991})\BibitemShut {NoStop}%
\bibitem [{\citenamefont {Keener}(2000)}]{Keener:18}%
  \BibitemOpen
  \bibfield  {author} {\bibinfo {author} {\bibfnamefont {J.~P.}\ \bibnamefont
  {Keener}},\ }\href@noop {} {\emph {\bibinfo {title} {Principles of Applied
  Mathematics: Transformation and Approximation}}}\ (\bibinfo  {publisher} {CRC
  Press},\ \bibinfo {year} {2000})\BibitemShut {NoStop}%
\bibitem [{\citenamefont {Happel}\ and\ \citenamefont
  {Brenner}(1965)}]{Happel:book}%
  \BibitemOpen
  \bibfield  {author} {\bibinfo {author} {\bibfnamefont {J.}~\bibnamefont
  {Happel}}\ and\ \bibinfo {author} {\bibfnamefont {H.}~\bibnamefont
  {Brenner}},\ }\href@noop {} {\emph {\bibinfo {title} {Low Reynolds Number
  Hydrodynamics}}}\ (\bibinfo  {publisher} {Prentice-Hall},\ \bibinfo {address}
  {Englewood Cliffs, N. J.},\ \bibinfo {year} {1965})\BibitemShut {NoStop}%
\bibitem [{\citenamefont {Stone}\ and\ \citenamefont
  {Samuel}(1996)}]{Stone:96}%
  \BibitemOpen
  \bibfield  {author} {\bibinfo {author} {\bibfnamefont {H.~A.}\ \bibnamefont
  {Stone}}\ and\ \bibinfo {author} {\bibfnamefont {A.~D.~T.}\ \bibnamefont
  {Samuel}},\ }\bibfield  {title} {\enquote {\bibinfo {title} {Propulsion of
  microorganisms by surface distortions},}\ }\href@noop {} {\bibfield
  {journal} {\bibinfo  {journal} {Phys. Rev. Lett.}\ }\textbf {\bibinfo
  {volume} {77}},\ \bibinfo {pages} {4102} (\bibinfo {year}
  {1996})}\BibitemShut {NoStop}%
\bibitem [{\citenamefont {De~Buyl}\ \emph {et~al.}(2013)\citenamefont
  {De~Buyl}, \citenamefont {Mikhailov},\ and\ \citenamefont {Kapral}}]{De:13}%
  \BibitemOpen
  \bibfield  {author} {\bibinfo {author} {\bibfnamefont {P.}~\bibnamefont
  {De~Buyl}}, \bibinfo {author} {\bibfnamefont {A.~S.}\ \bibnamefont
  {Mikhailov}}, \ and\ \bibinfo {author} {\bibfnamefont {R.}~\bibnamefont
  {Kapral}},\ }\bibfield  {title} {\enquote {\bibinfo {title} {Self-propulsion
  through symmetry breaking},}\ }\href@noop {} {\bibfield  {journal} {\bibinfo
  {journal} {EPL}\ }\textbf {\bibinfo {volume} {103}},\ \bibinfo {pages}
  {60009} (\bibinfo {year} {2013})}\BibitemShut {NoStop}%
\bibitem [{\citenamefont {Van~Bladel}(2007)}]{van:Book}%
  \BibitemOpen
  \bibfield  {author} {\bibinfo {author} {\bibfnamefont {J.~G.}\ \bibnamefont
  {Van~Bladel}},\ }\href@noop {} {\emph {\bibinfo {title} {Electromagnetic
  fields}}},\ Vol.~\bibinfo {volume} {19}\ (\bibinfo  {publisher} {John Wiley
  \& Sons},\ \bibinfo {year} {2007})\BibitemShut {NoStop}%
\end{thebibliography}%
\end{document}